\documentclass{aa}

\usepackage{txfonts}
\usepackage{graphicx}
\usepackage{natbib}
\bibpunct{(}{)}{;}{a}{}{,}

\begin{document}

\title{The ALHAMBRA survey\thanks{Based on observations collected at the German-Spanish Astronomical Center, Calar Alto, jointly operated by the Max-Planck-Institut f\"ur Astronomie (MPIA) at Heidelberg and the Instituto de Astrof\'{\i}sica de Andaluc\'{\i}a (IAA-CSIC).}: an empirical estimation of the cosmic variance for merger fraction studies based on close pairs}

\author{C.~L\'opez-Sanjuan\inst{1,}\thanks{\email{clsj@cefca.es}}
\and A.~J.~Cenarro              \inst{1}
\and C.~Hern\'andez-Monteagudo  \inst{1}
\and J.~Varela                  \inst{1}
\and A.~Molino			\inst{2}
\and P.~Arnalte-Mur		\inst{3}
\and B.~Ascaso			\inst{2}
\and F.~J.~Castander		\inst{4}
\and A.~Fern\'andez-Soto	\inst{5,6}
\and M.~Huertas-Company		\inst{7,8}
\and I.~M\'arquez		\inst{2}
\and V.~J.~Mart\'{\i}nez	\inst{6,9}
\and J.~Masegosa		\inst{2}
\and M.~Moles			\inst{1,2}
\and M.~Povi\'c			\inst{2}
\and J.~A.~L.~Aguerri		\inst{10,11}
\and E.~Alfaro			\inst{2}
\and N.~Ben\'{\i}tez		\inst{2}
\and T.~Broadhurst		\inst{12}
\and J.~Cabrera-Ca\~no		\inst{13}
\and J.~Cepa			\inst{10,11}
\and M.~Cervi\~no		\inst{2,10}
\and D.~Crist\'obal-Hornillos	\inst{1}
\and A.~Del Olmo		\inst{2}
\and R.~M.~Gonz\'alez Delgado	\inst{2}
\and C.~Husillos		\inst{2}
\and L.~Infante			\inst{14}
\and J.~Perea			\inst{2}
\and F.~Prada			\inst{2}
\and J.~M.~Quintana		\inst{2}
}

\institute{Centro de Estudios de F\'{\i}sica del Cosmos de Arag\'on, Plaza San Juan 1, 44001 Teruel, Spain
\and Instituto de Astrof\'{\i}sica de Andaluc\'{\i}a (IAA-CSIC), Glorieta de la astronom\'{\i}a s/n, 18008 Granada, Spain 
\and Institute for Computational Cosmology, Department of Physics, Durham University, South Road, Durham DH1 3LE, UK 
\and Institut de Ci\`encies de l'Espai (IEEC-CSIC), Facultat de Ci\'encies, Campus UAB, 08193 Bellaterra, Spain 
\and Instituto de F\'{\i}sica de Cantabria, Avenida de los Castros s/n, 39005 Santander, Spain 
\and Observatori Astron\`omic, Universitat de Val\`encia, C/ Catedr\'atico Jos\'e Beltr\'an 2, 46980 Paterna, Spain 
\and GEPI, Paris Observatory, 77 av. Denfert Rochereau, 75014 Paris, France 
\and University Denis Diderot, 4 Rue Thomas Mann, 75205 Paris, France 
\and Departament d'Astronomia i Astrof\'{\i}sica, Universitat de Val\`encia, 46100 Burjassot, Spain 
\and Instituto de Astrof\'{\i}sica de Canarias, V\'{\i}a L\'actea s/n, La Laguna, 38200 Tenerife, Spain 
\and Departamento de Astrof\'{\i}sica, Facultad de F\'{\i}sica, Universidad de la Laguna, 38200 La Laguna, Spain 
\and Department of Theoretical Physics, University of the Basque Country UPV/EHU, Bilbao, Spain 
\and Departamento de F\'{\i}sica At\'omica, Molecular y Nuclear, Facultad de F\'{\i}sica, Universidad de Sevilla, 41012 Sevilla, Spain 
\and Departamento de Astronom\'{\i}a y Astrof\'{\i}sica, Facultad de F\'{\i}sica, Pontificia Universidad Cat\'olica de Chile, 782-0436 Santiago, Chile 
}

\date{Received 12 August 2013 -- Accepted 17 January 2014}

\abstract
{}
{Our goal is to estimate empirically, for the first time, the cosmic variance that affects merger fraction studies based on close pairs.}
{We compute the merger fraction from photometric redshift close pairs with $10h^{-1}\ {\rm kpc} \leq r_{\rm p} \leq 50h^{-1}$ kpc and $\Delta v \leq 500$ km s$^{-1}$, and measure it in the 48 sub-fields of the ALHAMBRA survey. We study the distribution of the measured merger fractions, that follow a log-normal function, and estimate the cosmic variance $\sigma_v$ as the intrinsic dispersion of the observed distribution. We develop a maximum likelihood estimator to measure a reliable $\sigma_v$ and avoid the dispersion due to the observational errors (including the Poisson shot noise term).}
{The cosmic variance $\sigma_v$ of the merger fraction depends mainly on (i) the number density of the populations under study, both for the principal ($n_1$) and the companion ($n_2$) galaxy in the close pair, and (ii) the probed cosmic volume $V_c$. We find a significant dependence on neither the search radius used to define close companions, the redshift, nor the physical selection (luminosity or stellar mass) of the samples.}
{We have estimated from observations the cosmic variance that affects the measurement of the merger fraction by close pairs. We provide a parametrisation of the cosmic variance with $n_1$, $n_2$, and $V_c$, $\sigma_v \propto n_1^{-0.54} V_c^{-0.48} (n_2/n_1)^{-0.37}$. Thanks to this prescription, future merger fraction studies based on close pairs could account properly for the cosmic variance on their results.}


\titlerunning{The ALHAMBRA survey. An empirical estimation of the cosmic variance for merger fraction studies} 

\maketitle

\section{Introduction}\label{intro}
Our understanding of the formation and evolution of galaxies across cosmic time have been greatly improved in the last decade thanks to deep photometric and spectroscopic surveys. Some examples of these successful deep surveys are SDSS \citep[Sloan Digital Sky Survey,][]{sdssdr7}, GOODS \citep[Great Observatories Origins Deep Survey,][]{goods}, AEGIS \citep[All-Wavelength Extended Groth Strip International Survey,][]{aegis}, ELAIS \citep[European Large-Area ISO Survey,][]{elais}, COSMOS \citep[Cosmological Evolution Survey,][]{cosmos}, MGC \citep[Millennium Galaxy Catalogue,][]{mgc}, VVDS \citep[VIMOS VLT Deep Survey,][]{lefevre05,vvdsud}, DEEP \citep[Deep Extragalactic Evolutionary Probe,][]{deep2}, zCOSMOS \citep[][]{zcosmos10k}, GNS \citep[GOODS NICMOS Survey,][]{gns}, SXDS \citep[Subaru/XMM-Newton Deep Survey,][]{sxds}, or CANDELS \citep[Cosmic Assembly NIR Deep Extragalactic Legacy Survey,][]{candels,candels2}.

One fundamental uncertainty in any observational measurement derived from galaxy surveys is the {\it cosmic variance} ($\sigma_v$), arising from the underlying large-scale density fluctuations and leading to variances larger than those expected from simple Poisson statistics. The most efficient way to tackle with cosmic variance is split the survey in several independent areas in the sky. This minimises the sampling problem better than increase the volume in a wide contiguous field \citep[e.g.,][]{driver10}. However, observational constraints (depth vs area) lead to many existing surveys to have observational uncertainties dominated by the cosmic variance. Thus, a proper estimation of $\sigma_v$ is needed to fully describe the error budget in deep cosmological surveys.

The impact of the cosmic variance in a given survey and redshift range can be estimated using two basic methods: theoretically by analysing cosmological simulations \citep[e.g.,][]{somerville04,trenti08,stringer09,moster11}, or empirically by sampling a larger survey \citep[e.g.,][]{driver10}. Unfortunately, previous studies estimate only the cosmic variance affecting number density measurements, and do not tackle the impact of $\sigma_v$ in other important quantities as the merger fraction. Merger fraction studies based on close pair statistics measure the correlation of two galaxy populations at small scales ($\leq 100h^{-1}$ kpc), so the amplitude of the cosmic variance and its dependence on galaxy properties, probed volume, etc. should be different than those in number density studies. In the present paper we take advantage of the unique design, depth, and photometric redshift accuracy of the ALHAMBRA\footnote{http://alhambrasurvey.com} (Advanced, Large, Homogeneous Area, Medium-Band Redshift Astronomical) survey \citep{alhambra} to estimate empirically, for the first time, the cosmic variance that affect close pair studies. The ALHAMBRA survey has observed 8 separate regions of the northern sky, comprising 48 sub-fields of $\sim180$ arcmin$^{2}$ each that can be assumed as independent for our purposes. Thus, ALHAMBRA provides 48 measurements of the merger fraction across the sky. The intrinsic dispersion in the distribution of these merger fractions, that we characterise in the present paper, is an observational estimation of the cosmic variance $\sigma_v$.

The paper is organised as follows. In Sect.~\ref{data} we present the ALHAMBRA survey and its photometric redshifts, and in Sect.~\ref{metodo} we review the methodology to measure close pair merger fractions when photometric redshifts are used. We present our estimation and characterisation of the cosmic variance for close pair studies in Sect.~\ref{analysis}. In Sect.~\ref{conclusions} we summarise our work and present our conclusions. Throughout this paper we use a standard cosmology with $\Omega_{\rm m} = 0.3$, $\Omega_{\Lambda} = 0.7$, $H_{0}= 100h$ km s$^{-1}$ Mpc$^{-1}$, and $h = 0.7$. Magnitudes are given in the AB system.

\section{The ALHAMBRA survey}\label{data}

The ALHAMBRA survey provides a photometric data set over 20 contiguous, equal-width ($\sim$300\AA), non-overlapping, medium-band optical filters (3500\AA -- 9700\AA) plus 3 standard broad-band near-infrared (NIR) filters ($J$, $H$, and $K_{\rm s}$) over 8 different regions of the northern sky \citep{alhambra}. The survey has the aim of understanding the evolution of galaxies throughout cosmic time by sampling a large enough cosmological fraction of the universe, for which reliable spectral energy distributions (SEDs) and precise photometric redshifts ($z_{\rm p}$'s) are needed. The simulations of \citet{benitez09}, relating the image depth and $z_{\rm p}$'s accuracy to the number of filters, have demonstrated that the filter set chosen for ALHAMBRA can achieve a photometric redshift precision that is three times better than a classical $4 - 5$ optical broad-band filter set. The final survey parameters and scientific goals, as well as the technical properties of the filter set, were described by \citet{alhambra}. The survey has collected its data for the 20+3 optical-NIR filters in the 3.5m telescope at the Calar Alto observatory, using the wide-field camera LAICA (Large Area Imager for Calar Alto) in the optical and the OMEGA–2000 camera in the NIR. The full characterisation, description, and performance of the ALHAMBRA optical photometric system was presented in \citet{aparicio10}. A summary of the optical reduction can be found in Crist\'obal-Hornillos et al. (in prep.), while of the NIR reduction in \citet{cristobal09}.

\begin{figure}[t]
\centering
\resizebox{\hsize}{!}{\includegraphics{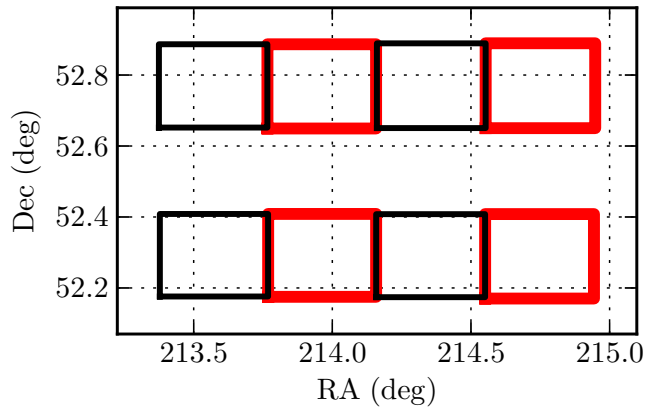}}
\caption{Schematic view of the ALHAMBRA field's geometry in the sky plane. We show the eight sub-fields (one per LAICA chip) of the field ALHAMBRA-6. The black and red squares mark the two LAICA pointings in this particular field. The geometry of the other seven fields is similar. [{\it A colour version of this plot is available in the electronic edition}].}
\label{alfield}
\end{figure}

The ALHAMBRA survey has observed 8 well-separated regions of the northern sky. The wide-field camera LAICA has four chips with a $15\arcmin \times 15\arcmin$ field-of-view each (0.22 arcsec/pixel). The separation between chips is also $15\arcmin$. Thus, each LAICA pointing provides four separated areas in the sky (black or red squares in Fig.~\ref{alfield}). Six ALHAMBRA regions comprise two LAICA pointings. In these cases, the pointings define two separate strips in the sky (Fig.~\ref{alfield}). In our study we assumed the four chips in each strip as independent sub-fields. The photometric calibration of the field ALHAMBRA-1 is currently on-ongoing, and the fields ALHAMBRA-4 and ALHAMBRA-5 comprise one pointing each \citep[see][for details]{molino13}. We summarise the properties of the 7 ALHAMBRA fields used in the present paper in Table~\ref{alhambra_fields_tab}. At the end, ALHAMBRA comprises 48 sub-fields of $\sim180$ arcmin$^2$, that we assumed independent, in which we measured the merger fraction following the methodology described in Sect.~\ref{metodo}. When we searched for close companions near the sub-field boundaries we did not consider the observed sources in the adjacent fields to keep the measurements independent. We prove the independence of the 48 ALHAMBRA sub-fields in Sect~\ref{sec7f}.

\begin{table*}
\caption{The ALHAMBRA survey fields}
\label{alhambra_fields_tab}
\begin{center}
\begin{tabular}{lcccc}
\hline\hline\noalign{\smallskip}
Field      &    Overlapping     &    RA    &    DEC     &    sub-fields / area \\
name       &      survey	&   (J2000) & (J2000)    &   (\# / deg$^2$)\\
\noalign{\smallskip}
\hline
\noalign{\smallskip}
ALHAMBRA-2  &  DEEP2	& 01 30 16.0	& +04 15 40   &  8 / 0.377	\\
ALHAMBRA-3  &  SDSS	& 09 16 20.0	& +46 02 20   &  8 / 0.404	\\
ALHAMBRA-4  &  COSMOS	& 10 00 00.0	& +02 05 11   &  4 / 0.203	\\
ALHAMBRA-5  &  GOODS-N	& 12 35 00.0	& +61 57 00   &  4 / 0.216	\\
ALHAMBRA-6  &  AEGIS	& 14 16 38.0	& +52 24 50   &  8 / 0.400	\\
ALHAMBRA-7  &  ELAIS-N1	& 16 12 10.0	& +54 30 15   &  8 / 0.406	\\
ALHAMBRA-8  &  SDSS	& 23 45 50.0	& +15 35 05   &  8 / 0.375	\\
Total	    &		&		&	      & 48 / 2.381\\
\noalign{\smallskip}
\hline
\end{tabular}
\end{center}
\end{table*}

\begin{figure}[t]
\centering
\resizebox{\hsize}{!}{\includegraphics{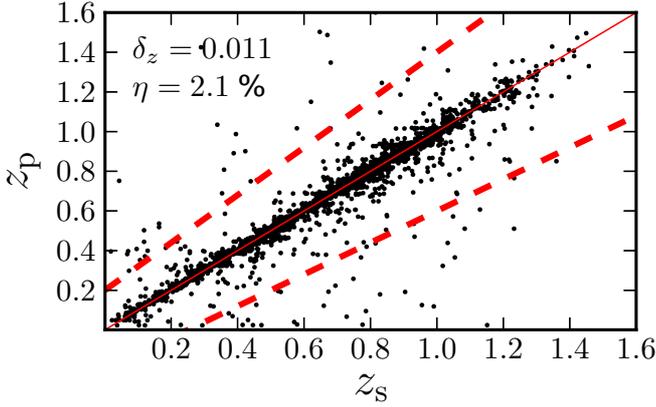}}
\caption{Photometric redshift ($z_{\rm p}$) versus spectroscopic redshift ($z_{\rm s}$) for the 3813 galaxies in the ALHAMBRA area with $i \leq 22.5$ and a measured $z_{\rm s}$. The solid line marks identity. The sources above and bellow the dashed lines are catastrophic outliers. The accuracy of the photometric redshifts ($\delta_z$) and the fraction of catastrophic outliers ($\eta$) are labelled in the panel. [{\it A colour version of this plot is available in the electronic edition}].}
\label{zpvszs}
\end{figure}

\subsection{Bayesian photometric redshifts in ALHAMBRA}
We rely on the ALHAMBRA photometric redshifts to compute the merger fraction (Sect.~\ref{metodo}). The photometric redshifts used all over present paper are fully presented and tested in \citet{molino13}, and we summarise their principal characteristics below.

The ALHAMBRA $z_{\rm p}$'s were estimated with BPZ2.0, a new version of BPZ \citep{benitez00}. BPZ is a SED-fitting method based in a Bayesian inference where a maximum likelihood is weighted by a prior probability. The library of 11 SEDs (4 ellipticals, 1 lenticular, 2 spirals, and 4 starbursts) and the prior probabilities used by BPZ2.0 in ALHAMBRA are detailed in Ben\'{\i}tez (in prep.). The ALHAMBRA photometry used to compute the photometric redshifts is PSF-matched aperture-corrected and based on isophotal magnitudes. In addition, a recalibration of the zero point of the images was performed to enhance the accuracy of the $z_{\rm p}$'s. Sources were detected in a synthetic $F814W$ filter image, noted $i$ in the following, defined to resemble the HST/$F814W$ filter. The areas of the images affected by bright stars, as well as those with lower exposure times (e.g., the edges of the images), were masked following \citet{arnaltemur13}. The total area covered by the ALHAMBRA survey after masking is 2.38 deg$^{2}$. Finally, a statistical star/galaxy separation is encoded in the variable \texttt{Stellar\_Flag} of the ALHAMBRA catalogues, and throughout present paper we keep as galaxies those ALHAMBRA sources with $\texttt{Stellar\_Flag} \leq 0.5$.

The photometric redshift accuracy, estimated by comparison with spectroscopic redshifts ($z_{\rm s}$'s), is $\delta_z = 0.0108$ at $i \leq 22.5$ with a fraction of catastrophic outliers of $\eta = 2.1$\%. The variable $\delta_z$ is the normalized median absolute deviation of the photometric versus spectroscopic redshift distribution \citep{ilbert06,eazy},
\begin{equation}
\delta_z = 1.48 \times {\rm median}\,\bigg( \frac{|z_{\rm p} - z_{\rm s}|}{1 + z_{\rm s}} \bigg).
\end{equation} 
The variable $\eta$ is defined as the fraction of galaxies with $|z_{\rm p} - z_{\rm s}|/(1 + z_{\rm s}) > 0.2$. We illustrate the high quality of the ALHAMBRA photometric redshifts in Fig.~\ref{zpvszs}. We refer to \citet{molino13} for a more detailed discussion.

\begin{figure}[t]
\centering
\resizebox{\hsize}{!}{\includegraphics{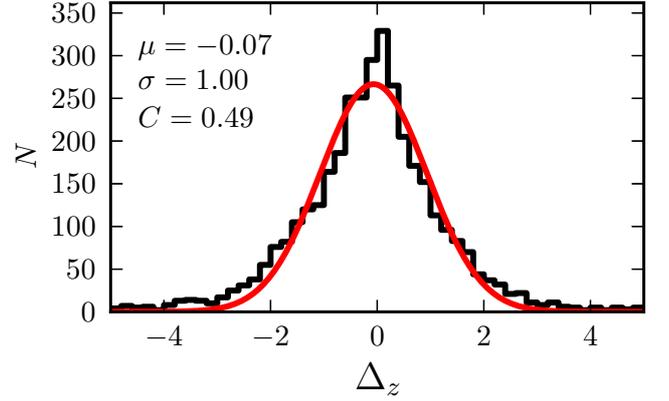}}
\caption{Distribution of the variable $\Delta_z$ for the 3813 galaxies in the ALHAMBRA area with $i \leq 22.5$ and a measured spectroscopic redshift. The red line is the best least-squares fit of a Gaussian function to the data. The median, dispersion and the factor $C$ derived from the fit are labelled in the panel. [{\it A colour version of this plot is available in the electronic edition}].}
\label{deltaz_dist}
\end{figure}

The \texttt{odds} quality parameter, noted $\mathcal{O}$, is a proxy for the photometric redshift accuracy of the sources and is also provided by BPZ2.0. The \texttt{odds} is defined as the redshift probability enclosed on a $\pm K(1+z)$ region around the main peak in the probability distribution function (PDF) of the source, where the constant $K$ is specific for each photometric survey. \citet{molino13} find that $K = 0.0125$ is the optimal value for the ALHAMBRA survey. The parameter $\mathcal{O} \in [0,1]$ is related with the confidence of the $z_{\rm p}$, making possible to derive high quality samples with better accuracy and lower rate of catastrophic outliers. For example, a $\mathcal{O} \geq 0.5$ selection for $i \leq 22.5$ galaxies yields $\delta_z = 0.0094$ and $\eta = 1$\%, while $\delta_z = 0.0061$ and $\eta = 0.8$\% for $\mathcal{O} \geq 0.9$ \citep[see][for further details]{molino13}. We explore the optimal \texttt{odds} selection in ALHAMBRA for close pair studies in Sect.~\ref{optimal}.

Reliable photometric redshift errors ($\sigma_{z_{\rm p}}$) are needed to compute the merger fraction in photometric samples (Sect.~\ref{metodo}). In addition to the $z_{\rm p}$, we have the $z_{\sigma}^{+}$ and $z_{\sigma}^{-}$ of each source, defined as the redshifts that enclose 68\% of the PDF of the source. We estimated the photometric redshift error of each individual source as $\sigma_{z_{\rm p}} = C \times (z_{\sigma}^{+} - z_{\sigma}^{-})$. The constant $C$ is estimated from the distribution of the variable 
\begin{equation}
\Delta_z = \frac{z_{\rm p} - z_{\rm s}}{\sigma_{z_{\rm p}}} = \frac{z_{\rm p} - z_{\rm s}}{C \times (z_{\sigma}^{+} - z_{\sigma}^{-})}.
\end{equation}
The variable $\Delta_z$ should be normally distributed with zero mean and unit variance if the $\sigma_{z_{\rm p}}$'s from ALHAMBRA are a good descriptor of the $z_{\rm p}$'s accuracy \citep[e.g.,][]{ilbert09,carrasco13}. We find that $\Delta_z$ is described well by a normal function when $C = 0.49$ (Fig.~\ref{deltaz_dist}, see also \citealt{molino13}). Note that, with the definition of $z_{\sigma}^{+}$ and $z_{\sigma}^{-}$, $C = 0.5$ was expected. This result also implies that the Gaussian approximation of the PDF assumed in the estimation of the merger fraction (Setc.~\ref{metodo}) is statistically valid, even if the actual PDF of the individual sources could be multimodal and/or asymmetric at faint magnitudes. We estimated $C$ for different $i$-band magnitudes and \texttt{odds} selections, finding that the $C$ values are consistent with the global one within $\pm 0.1$. Thus, we conclude that $\sigma_{z_{\rm p}}$ provides a reliable photometric redshift error for every ALHAMBRA source.

\subsection{Sample selection}
Throughout present paper we focus our analysis in the galaxies of the ALHAMBRA first data release\footnote{http://cloud.iaa.es/alhambra/}. This catalogue comprises $\sim500$k sources and is complete ($5\sigma$, $3\arcsec$ aperture) for $i \leq 24.5$ galaxies \citep{molino13}. We explored different apparent luminosity sub-samples from $i \leq 23$ to $i \leq 20$. That ensures excellent photometric redshifts and provides reliable merger fraction measurements (Sect.~\ref{optimal}), because the PDFs of $i \leq 23$ sources are defined well by a single Gaussian peak \citep[][]{molino13}. In Sect.~\ref{secpop} we also study the cosmic variance in luminosity- and stellar mass-selected samples. The $B-$band luminosities and the stellar masses of the ALHAMBRA sources were also provided by BPZ2.0 and are included in the ALHAMBRA catalogue \citep[see][for further details]{molino13}. The mass-to-light ratios from \citet{taylor11} and a \citet{chabrier03} initial mass function were assumed in the estimation of the stellar masses.

\section{Measuring of the merger fraction in photometric samples}\label{metodo}
The linear distance between two sources can be obtained from their projected separation, $r_{\rm p} = \phi\,d_A(z_1)$, and their rest-frame relative velocity along the line of sight, $\Delta v = {c\, |z_2 - z_1|}/(1+z_1)$, where $z_1$ and $z_2$ are the redshift of the principal (more luminous/massive galaxy in the pair) and the companion galaxy, respectively; $\phi$ is the angular separation, in arcsec, of the two galaxies on the sky plane; and $d_A(z)$ is the angular diameter distance, in kpc arcsec$^{-1}$, at redshift $z$. Two galaxies are defined as a close pair if $r_{\rm p}^{\rm min} \leq r_{\rm p} \leq r_{\rm p}^{\rm max}$ and $\Delta v \leq \Delta v^{\rm max}$. The PSF of the ALHAMBRA ground-based images is $\lesssim 1.4\arcsec$ (median seeing of $\sim1\arcsec$), which corresponds to $7.6h^{-1}$ kpc in our cosmology at $z = 0.9$. To ensure well de-blended sources and to minimise colour contamination, we fixed $r_{\rm p}^{\rm min}$ to $10h^{-1}$ kpc ($\phi > 1.8\arcsec$ at $z < 0.9$). We left $r_{\rm p}^{\rm max} \leq 50h^{-1}$ kpc as a free parameter and estimate its optimal value in Sect.~\ref{optimal}. Finally, we set $\Delta v^{\rm max} = 500$ km s$^{-1}$ following spectroscopic studies \citep[e.g.,][]{patton00,lin08}. With the previous constraints 50\%-70\% of the selected close pairs will finally merge \citep{patton08,bell06,jian12}.

To compute close pairs we defined a principal and a companion sample. The principal sample comprises the more luminous or massive galaxy of the pair, and we looked for those galaxies in the companion sample that fulfil the close pair criterion for each galaxy of the principal sample. If one principal galaxy has more than one close companion, we took each possible pair separately (i.e., if the companion galaxies B and C are close to the principal galaxy A, we study the pairs A-B and A-C as independent). In addition, through present paper we do not impose any luminosity or mass difference between the galaxies in the close pair unless noted otherwise.

With the previous definitions the merger fraction is
\begin{equation}
f_{\rm m}\ = \frac{N_{\rm p}}{N_{1}},\label{ncspec}
\end{equation}
where $N_1$ is the number of sources in the principal sample and $N_{\rm p}$ the number of close pairs. This definition applies to spectroscopic volume-limited samples, but we rely on photometric redshifts to compute $f_{\rm m}$ in ALHAMBRA. In a previous work, \citet{clsj10pargoods} develop a statistical method to obtain reliable merger fractions from photometric redshift catalogues as those from the ALHAMBRA survey. This methodology has been tested with the MGC \citep{clsj10pargoods} and the VVDS \citep{clsj12sizecos} spectroscopic surveys, and successfully applied in the GOODS-South \citep{clsj10pargoods} and the COSMOS fields \citep{clsj12sizecos}. We recall the main points of this methodology below and we explore how to apply it optimally over the ALHAMBRA data in Sect.~\ref{optimal}.

We used the following procedure to define a close pair system in our photometric catalogue \citep[see][for details]{clsj10pargoods}: first we search for close spatial companions of a principal galaxy, with redshift $z_1$ and uncertainty $\sigma_{z_1}$, assuming that the galaxy is located at $z_1 - 2\sigma_{z_1}$. This defines the maximum $\phi$ possible for a given $r_{\rm p}^{\rm max}$ in the first instance. If we find a companion galaxy with redshift $z_2$ and uncertainty $\sigma_{z_2}$ at $r_{\rm p} \leq r_{\rm p}^{\rm max}$, we study both galaxies in redshift space. For convenience, we assume below that every principal galaxy has, at most, one close companion. In this case, our two galaxies could be a close pair in the redshift range
\begin{equation}
[z^{-},z^{+}] = [z_1 - 2\sigma_{z_1}, z_1 + 2\sigma_{z_1}] \cap [z_2- 2\sigma_{z_2}, z_2 + 2\sigma_{z_2}].
\end{equation}
Because of variation in the range $[z^{-},z^{+}]$ of the function $d_A(z)$, a sky pair at $z_1 - 2\sigma_{z_1}$ might not be a pair at $z_1 + 2\sigma_{z_1}$. We thus impose the condition $r_{\rm p}^{\rm min} \leq r_{\rm p} \leq r_{\rm p}^{\rm max}$ at all $z \in [z^{-},z^{+}]$, and redefine this redshift interval if the sky pair condition is not satisfied at every redshift. After this, our two galaxies define the close pair system $k$ in the redshift interval $[z^{-}_k,z^{+}_k]$, where the index $k$ covers all the close pair systems in the sample.

The next step is to define the number of pairs associated to each close pair system $k$. For this, and because all our sources have a photometric redshift, we suppose in the following that a galaxy $i$ in whatever sample is described in redshift space by a Gaussian probability distribution, 
\begin{equation}
P_i\,(z_i\,|\,z_{{\rm p},i},\sigma_{z_{{\rm p},i}}) = \frac{1}{\sqrt{2\pi}\sigma_{z_{{\rm p},i}}}\exp\bigg[-\frac{(z_i-z_{{\rm p},i})^2}{2\sigma_{z_{{\rm p},i}}^2}\bigg]\label{zgauss}.
\end{equation}

With the previous distribution we are able to treat statistically all the available information in redshift space and define the number of pairs at redshift $z_1$ in system $k$ as
\begin{equation}
\nu_{k}\,(z_1) = {\rm C}_k\, P_1 (z_1\, |\, z_{{\rm p},1},\sigma_{z_{{\rm p},1}}) \int_{z_{\rm m}^{-}}^{z_{\rm m}^{+}} P_2 (z_2\, |\, z_{{\rm p},2},\sigma_{z_{{\rm p},2}})\, {\rm d}z_2,\label{nuj}
\end{equation}
where $z_1 \in [z^{-}_k,z^{+}_k]$, the integration limits are
\begin{eqnarray}
z_{\rm m}^{-} = z_1(1-\Delta v^{\rm max}/c) - \Delta v^{\rm max}/c,\\
z_{\rm m}^{+} = z_1(1+\Delta v^{\rm max}/c) + \Delta v^{\rm max}/c,
\end{eqnarray}
the subindex 1 [2] refers to the principal [companion] galaxy in the system $k$, and the constant ${\rm C}_k$ normalises the function to the total number of pairs in the interest range,
\begin{equation}
2 N_{\rm p}^k = \int_{z_k^{-}}^{z_k^{+}} P_1 (z_1\, |\, z_{{\rm p},1},\sigma_{z_{{\rm p},1}})\, {\rm d}z_1  + \!\! \int_{z_k^{-}}^{z_k^{+}} P_2 (z_2\, |\, z_{{\rm p},2},\sigma_{z_{{\rm p},2}})\, {\rm d}z_2.
\end{equation}
Note that $\nu_k = 0$ if $z_1 < z_k^-$ or  $z_1 > z_k^+$. The function $\nu_k$ tells us how the number of pairs in the system $k$, noted $N_{\rm p}^k$, are distributed in redshift space. The integral in Eq.~(\ref{nuj}) spans those redshifts in which the companion galaxy has $\Delta v \leq \Delta v^{\rm max}$ for a given redshift of the principal galaxy. This translates to $z_{\rm m}^{+} - z_{\rm m}^{-} \sim 0.005$ in our redshift range of interest.

With the previous definitions, the merger fraction in the interval $z_{\rm r} = [z_{\rm min}, z_{\rm max})$ is
\begin{equation}
f_{{\rm m}} = \frac{\sum_k \int_{z_{\rm min}}^{z_{\rm max}}{\nu_k\,(z_1)}\, {\rm d}z_1}{\sum_i \int_{z_{\rm min}}^{z_{\rm max}} P_i\, (z_i\,|\,z_{{\rm p},i},\sigma_{z_{{\rm p},i}})\, {\rm d}z_i}.\label{ncphot}
\end{equation}
If we integrate over the whole redshift space, $z_{\rm r} = [0,\infty)$, Eq.~(\ref{ncphot}) becomes
\begin{equation}
f_{{\rm m}} = \frac{\sum_k N_{\rm p}^k}{N_1},\label{ncphot2}
\end{equation}
where $\sum_k N_{\rm p}^k$ is analogous to $N_{\rm p}$ in Eq.~(\ref{ncspec}). In order to estimate the observational error of $f_{{\rm m}}$, noted $\sigma_{f}$, we used the jackknife technique \citep{efron82}. We computed partial standard deviations, $\delta_k$, for each system $k$ by taking the difference between the measured $f_{{\rm m}}$ and the same quantity with the $k$th pair removed for the sample, $f_{{\rm m}}^k$, such that $\delta_k = f_{{\rm m}} - f_{{\rm m}}^k$. For a redshift range with $N_{\rm p}$ systems, the variance is given by $\sigma_{f}^2 = [(N_{\rm p}-1) \sum_k \delta_k^2]/N_{\rm p}$.

\subsection{Border effects in redshift and in the sky plane}\label{border}
When we search for a primary source's companion, we define a volume in the sky plane-redshift space. If the primary source is near the boundaries of the survey, a fraction of the search volume lies outside of the effective volume of the survey. \citet{clsj10pargoods} find that border effects in the sky plane are representative (i.e., $1\sigma$ discrepancy) only at $r_{\rm p}^{\rm max} \gtrsim 70h^{-1}$ kpc. Thus, we restricted the search radius in our study to $r_{\rm p}^{\rm max} \leq 50h^{-1}$ kpc.

We avoid the incompleteness in redshift space by including in the samples not only the sources inside the redshift range $[z_{\rm min}, z_{\rm max})$ under study, but also those sources with either $z_{{\rm p},i} + 2\sigma_{z_{{\rm p},i}} \geq z_{\rm min}$ or $z_{{\rm p},i} - 2\sigma_{z_{{\rm p},i}} < z_{\rm max}$. 

\subsection{The merger rate}\label{secmr}
The final goal of merger studies is the estimation of the merger rate $R_{\rm m}$, defined as the number of mergers per galaxy and Gyr$^{-1}$. The merger rate is computed from the merger fraction by close pairs as
\begin{equation}
R_{\rm m} = \frac{C_{\rm m}}{T_{\rm m}}\,f_{\rm m},
\end{equation}
where $C_{\rm m}$ is the fraction of the observed close pairs than finally merge after a merger time scale $T_{\rm m}$. The merger time scale and the merger probability $C_{\rm m}$ should be estimated from simulations \citep[e.g.,][]{kit08,lotz10gas,lotz10t,lin10,jian12,moreno13}. On the one hand, $T_{\rm m}$ depends mainly on the search radius $r_{\rm p}^{\rm max}$, the stellar mass of the principal galaxy, and the mass ratio between the galaxies in the pair, with a mild dependence on redshift and environment \citep{jian12}. On the other hand, $C_{\rm m}$ depends mainly on $r_{\rm p}^{\rm max}$ and environment, with a mild dependence on both redshift and the mass ratio between the galaxies in the pair \citep{jian12}. Despite of the efforts in the literature to estimate both $T_{\rm m}$ and $C_{\rm m}$, different cosmological and galaxy formation models provide different values within a factor of two--three \citep[e.g.,][]{hopkins10mer}. To avoid model-dependent results, in the present paper we focus therefore in the cosmic variance of the observational merger fraction $f_{\rm m}$.

\section{Estimation of the cosmic variance for merger fraction studies}\label{analysis}

\subsection{Theoretical background}\label{theory}
In this section we recall the theoretical background and define the basic variables involved in the cosmic variance definition and characterisation. The {\it relative cosmic variance} ($\sigma_v$) arises from the underlying large-scale density fluctuations and lead to variances larger than those expected from simple Poisson statistics. Following \citet{somerville04} and \citet{moster11}, the mean $\langle N \rangle$ and the variance $\langle N^2 \rangle - \langle N \rangle^2$ in the distribution of galaxies are given by the first and second moments of the probability distribution $P_N(V_c)$, which describes the probability of counting $N$ objects within a volume $V_c$. The relative cosmic variance is defined as
\begin{equation}
\sigma_v^2 = \frac{\langle N^2 \rangle - \langle N \rangle^2}{\langle N \rangle^2} - \frac{1}{\langle N \rangle}.\label{cosvarteo}
\end{equation}
The second term represents the correction for the Poisson shot noise. The second moment of the object counts is
\begin{equation}
\langle N^2 \rangle = \langle N \rangle^2 + \langle N \rangle + \frac{\langle N \rangle^2}{V_c^2} \int_{V_c} \xi(|{r}_{\rm a} - r_{\rm b}|)\,{\rm d}V_{c,{\rm a}}\,{\rm d}V_{c,{\rm b}},
\end{equation}
where $\xi$ is the two-point correlation function of the sample under study \citep{peebles80}. Combining this with Eq.~(\ref{cosvarteo}), the relative cosmic variance can be written as
\begin{equation}
\sigma_v^2 = \frac{1}{V_c^2}\,\int_{V_c} \xi(|{\bf r}_{\rm a} - {\bf r}_{\rm b}|)\,{\rm d}V_{c,{\rm a}}\,{\rm d}V_{c,{\rm b}}.\label{cosvarteoxi}
\end{equation}
Thus, the cosmic variance of a given sample depends on the correlation function of that population. We can approximate the galaxy correlation function in Eq.~(\ref{cosvarteoxi}) by the linear theory correlation function for dark matter $\xi_{\rm dm}$, $\xi = b^2\,\xi_{\rm dm}$, where $b$ is the galaxy bias. The bias at a fixed scale depends mainly on both redshift and the selection of the sample under study. With this definition of the correlation function we find that
\begin{equation}
\sigma_v \propto \frac{b}{V_c^{1 - \alpha}},
\end{equation}
where the power law index $\alpha$ takes into account the extra volume dependence from the integral of the correlation function $\xi_{\rm dm}$ in Eq.~(\ref{cosvarteoxi}).

The bias of a particular population is usually measured from the analysis of the correlation function and is well established that the bias increases with luminosity and stellar mass \citep[see][and references therein]{zehavi11,coupon12,marulli13,arnaltemur13}. The estimation of the bias is a laborious task, so we decided to use the redshift and the number density $n$ of the population under study instead of the bias to characterise the cosmic variance. The number density is an observational quantity that decreases with the increase of the luminosity and the mass selection, so a $b \propto n^{-\beta}$ relation is expected. This inverse dependence is indeed suggested by \citet{nuza13} results. 

In summary, we expect
\begin{equation}
\sigma_v \propto \frac{b}{V_c^{1 - \alpha}} \propto \frac{z^{\gamma}}{n^{\beta}\,V_c^{1 - \alpha}}.\label{cosvarteofin}
\end{equation}
This equation shows that the number density of galaxies, the redshift, and the cosmic volume can be assumed as independent variables in the cosmic variance parametrisation. Equation~(\ref{cosvarteofin}) and the deduction above apply to the cosmic variance in the number of galaxies. We are interested on the cosmic variance of the merger fraction by close pairs instead, so a dependence on $V_c$, redshift, and the number density of the two populations under study, noted $n_1$ for principal galaxies and $n_2$ for the companion galaxies, is expected. We used therefore this four variables ($n_1$, $n_2$, $z$, and $V_c$) to characterise the cosmic variance in close pair studies (Sect.~\ref{sigv}).

The power-law indices in Eq.~(\ref{cosvarteofin}) could be different for luminosity- and mass-selected samples, as well as for flux-limited samples. In the present paper we use flux-limited samples selected in the $i$ band to characterise the cosmic variance. This choice has several benefits, since we have a well controlled selection function, a better understanding of the photometric redshifts and their errors, and we have access to larger samples at lower redshift that in the luminosity and the stellar mass cases. That improves the statistics and increases the useful redshift range. At the end, future studies will be interested on the cosmic variance in physically selected samples (i.e., luminosity or stellar mass). Thus, in Sect.~\ref{secpop} we compare the results from the flux-limited $i-$band samples with the actual cosmic variance measured in physically selected samples.

[C3] Finally, we set the definition of the number density $n$. In the present paper the number density of a given population is the {\it cosmic average number density} of that population. For example, if we are studying the merger fraction in a volume dominated by a cluster, we should not use the number density in that volume, but the number density derived from a general luminosity or mass function work instead. Thanks to the 48 sub-fields in ALHAMBRA we have direct access to the average number densities of the populations under study (Sect.~\ref{secn1}).

\begin{figure}[t]
\centering
\resizebox{\hsize}{!}{\includegraphics{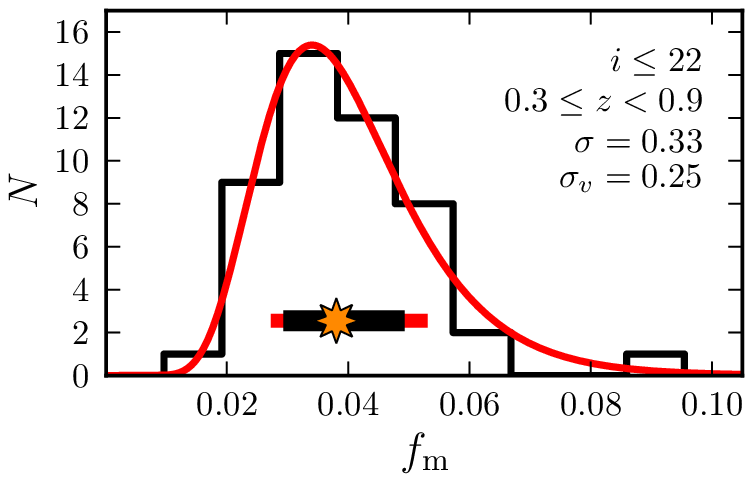}}
\resizebox{\hsize}{!}{\includegraphics{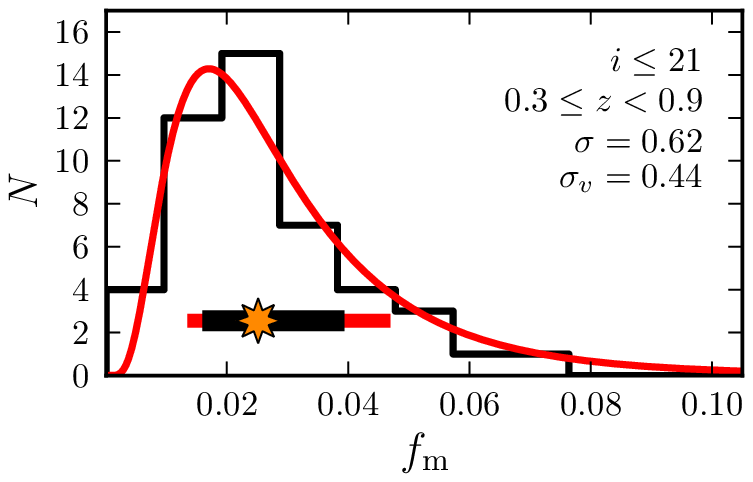}}
\caption{Distribution of the merger fraction $f_{\rm m}$ for $i \leq 22$ ({\it top panel}) and $i \leq 21$ ({\it bottom panel}) galaxies in the 48 ALHAMBRA sub-fields, measured from close pairs with $10h^{-1}\ {\rm kpc} \leq r_{\rm p} \leq 30h^{-1}$ kpc at $0.3 \leq z < 0.9$. In each panel, the red solid line is the best least-squares fit of a log-normal function to the data. The star and the red bar mark the median and the 68\% confidence interval of the fit, respectively. The black bar marks the confidence interval from the maximum likelihood analysis of the data and is our measurement of the cosmic variance $\sigma_v$. [{\it A colour version of this plot is available in the electronic edition}].}
\label{ff_lognormal}
\end{figure}

\subsection{Distribution of the merger fraction and $\sigma_{v}$ estimation}\label{ffdist}
In this section we explore which statistical distribution reproduces better the observed merger fractions and how to measure reliably the cosmic variance $\sigma_{v}$. As representative examples, we show in Fig.~\ref{ff_lognormal} the distributions of the merger fraction $f_{\rm m}$ in the 48 ALHAMBRA sub-fields for $i \leq 22$ and $i \leq 21$ galaxies. The merger fraction was measured from close pairs with $10h^{-1}\ {\rm kpc} \leq r_{\rm p} \leq 30h^{-1}$ kpc. Unless noted otherwise, in the following the principal and the companion samples comprise the same galaxies. We find that the observed distributions are not Gaussian, but follow a log-normal distribution instead,
\begin{equation}
P_{LN}\,(f_{\rm m}\,|\,\mu, \sigma) = \frac{1}{\sqrt{2 \pi}\,\sigma f_{\rm m}}\,{\rm exp}\,\bigg[-\frac{(\ln f_{\rm m} - \mu)^2}{2 \sigma^2}\bigg]\,,\label{Plog}
\end{equation}
where $\mu$ and $\sigma$ are the median and the dispersion of a Gaussian function in log-space $f'_{\rm m} = \ln f_{\rm m}$. This is,
\begin{equation}
P_{G}\,(f'_{\rm m}\,|\,\mu, \sigma) = \frac{1}{\sqrt{2 \pi}\,\sigma}\,{\rm exp}\,\bigg[-\frac{(f'_{\rm m} - \mu)^2}{2 \sigma^2}\bigg]\,.\label{Pg}
\end{equation}
The 68\% confidence interval of the log-normal distribution is $[{\rm e}^{\mu}{\rm e}^{-\sigma}, {\rm e}^{\mu}{\rm e}^{\sigma}]$. This functional distribution was expected for two reasons. First, the merger fraction can not be negative, implying an asymmetric distribution \citep{cameron11}. Second, the distribution of overdense structures in the universe is log-normal \citep[e.g.,][]{coles91,delatorre10,kovac10} and the merger fraction increases with density \citep[][]{lin10,deravel11,pawel12}. We checked that the merger fraction follows a log-normal distribution in all the samples explored in the present paper.

The variable $\sigma$ encodes the relevant information about the dispersion in the merger fraction distribution, including the dispersion due to the cosmic variance. The study of the median value of the merger fraction in ALHAMBRA, estimated as ${\rm e}^{\mu}$, and its dependence on $z$, stellar mass, or colour, is beyond the scope of the present paper and we will address this issue in a future work. 

\begin{figure}[t]
\centering
\resizebox{\hsize}{!}{\includegraphics{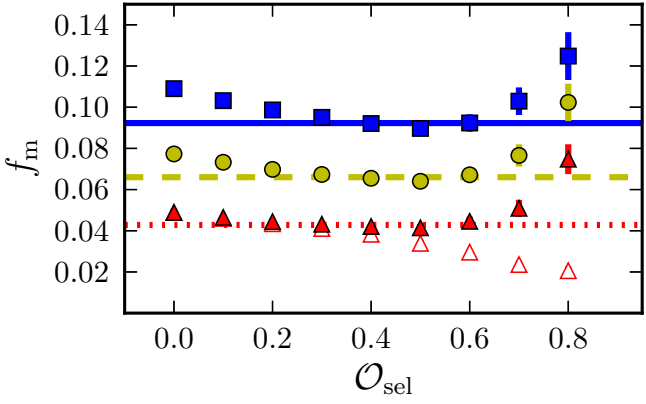}}
\caption{Merger fraction $f_{\rm m}$ as a function of the \texttt{odds} selection $\mathcal{O}_{\rm sel}$ for $i \leq 22.5$ galaxies at $0.3 \leq z < 0.9$. The filled triangles, circles, and squares are for $r_{\rm p}^{\rm max} = 30, 40$, and 50$h^{-1}$ kpc close pairs, respectively. The open triangles are the observed merger fractions for $r_{\rm p}^{\rm max} = 30h^{-1}$ kpc to illustrate the selection correction from Eq.~(\ref{fmal}). In several cases the error bars are smaller than the points. The dotted, dashed, and solid lines mark the average $f_{\rm m}$ at $0.3 \leq \mathcal{O_{\rm sel}} \leq 0.6$ for $r_{\rm p}^{\rm max} = 30, 40$, and 50$h^{-1}$ kpc close pairs, respectively. [{\it A colour version of this plot is available in the electronic edition}].}
\label{fmvsodds}
\end{figure}

A best least-squares fit with a log-normal function to the distributions in Fig.~\ref{ff_lognormal} shows that $\sigma$ increases with the apparent brightness, from $\sigma = 0.33$ for $i \leq 22$ galaxies to $\sigma = 0.62$ for $i \leq 21$ galaxies. However, the origin of the observed $\sigma$ is twofold: (i) the intrinsic dispersion due to the cosmic variance $\sigma_v$ (i.e., the field-to-field variation in the merger fraction because of the clustering of the galaxies), and (ii) the dispersion due to the observational errors $\sigma_{\rm o}$ (i.e., the uncertainty in the measurement of the merger fraction in a given field, including the Poisson shot noise term). Thus, the dispersion $\sigma$ reported in Fig.~\ref{ff_lognormal} is an upper limit for the actual cosmic variance $\sigma_v$. We deal with this limitation applying a maximum likelihood estimator (MLE) to the observed distributions. In Appendix~\ref{mlmethod} we develop a MLE that estimates the more probable values of $\mu$ and $\sigma_v$, assuming that the merger fraction follows a Gaussian distribution in log-space (Eq.~[\ref{Pg}]) that is affected by known observational errors $\sigma_{\rm o}$. We prove that the MLE provides an unbiased estimation of $\mu$ and $\sigma_v$, as well as reliable uncertainties of these parameters. Applying the MLE to the distributions in Fig.~\ref{ff_lognormal}, we find than $\sigma_v$ is lower than $\sigma$, as anticipated, and that the cosmic variance increases with the apparent brightness from $\sigma_v = 0.25 \pm 0.04$ for $i \leq 22$ galaxies to $\sigma_v = 0.44 \pm 0.08$ for $i \leq 21$ galaxies.

We constraint the dependence of $\sigma_v$ on the number density of the populations under study in Sects.~\ref{secn1} and \ref{secn2}, on the probed cosmic volume in Sect.~\ref{secvc}, and on redshift on Sect.~\ref{secz}. That provides a complete description of the cosmic variance for merger fraction studies. We stress that our definition of $\sigma_v$ differs from the classical definition of the relative cosmic variance presented in Sect.~\ref{theory}, which is equivalent to ${\rm e}^{\sigma_v}$. However, $\sigma_v$ encodes the relevant information needed to estimate the intrinsic dispersion in the measurement of the merger fraction due to the clustering of galaxies. 

\begin{figure}[t]
\centering
\resizebox{\hsize}{!}{\includegraphics{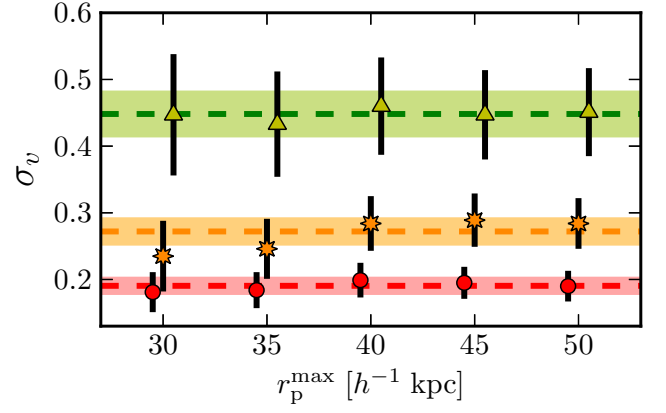}}
\caption{Cosmic variance $\sigma_v$ as a function of $r_{\rm p}^{\rm max}$ for $i \leq 22.5, 21.5$, and $21$ galaxies at $0.3 \leq z < 0.9$ (circles, stars, and triangles, respectively). The horizontal lines mark the error-weighted average of the cosmic variance in each case, and the coloured areas their 68\% confidence intervals. [{\it A colour version of this plot is available in the electronic edition}].}
\label{sigv_vs_rp}
\end{figure}

\begin{table}
\caption{Cosmic variance $\sigma_v$ as a function of the search radius $r_{\rm p}^{\rm max}$ for $\mathcal{O} \geq \mathcal{O}_{\rm sel} = 0.3$ galaxies at $0.3 \leq z < 0.9$}
\label{sigv_rp_tab}
\begin{center}
\begin{tabular}{lcccc}
\hline\hline\noalign{\smallskip}
$r_{\rm p}^{\rm max}$ &        $\sigma_v$      &      $\sigma_v$    & $\sigma_v$  \\
  ($h^{-1}$ kpc)      &     ($i \leq 22.5$)    &  ($i \leq 21.5$)   &  ($i \leq 21.0$)  \\
\noalign{\smallskip}
\hline
\noalign{\smallskip}
30 	& $0.181 \pm 0.030$ & $0.235 \pm 0.053$ & $0.447 \pm 0.091$\\
35 	& $0.184 \pm 0.027$ & $0.246 \pm 0.045$ & $0.433 \pm 0.079$\\
40 	& $0.199 \pm 0.026$ & $0.284 \pm 0.041$ & $0.460 \pm 0.073$\\
45 	& $0.195 \pm 0.024$ & $0.289 \pm 0.040$ & $0.447 \pm 0.067$\\
50 	& $0.190 \pm 0.023$ & $0.284 \pm 0.038$ & $0.451 \pm 0.066$\\
\noalign{\smallskip}
\hline
\noalign{\smallskip}
Average & $0.190 \pm 0.011$ & $0.272 \pm 0.019$ & $0.448 \pm 0.033$\\
\noalign{\smallskip}
\hline
\end{tabular}
\end{center}
\end{table}

\subsection{Optimal estimation of $\sigma_{v}$ in the ALHAMBRA survey}\label{optimal}

In the previous section we have defined the methodology to compute the cosmic variance from the observed distribution of the merger fraction. However, as shown by \citet{clsj10pargoods}, to avoid projection effects we need a galaxy sample with either small photometric redshift errors or a large fraction of spectroscopic redshifts. In the present study we did not use information from spectroscopic redshifts, so we should check that the photometric redshifts in ALHAMBRA are good enough for our purposes. A natural way to select excellent $z_{\rm p}$'s in ALHAMBRA is by a selection in the \texttt{odds} parameter. On the one hand, this selection increases the accuracy of the photometric redshifts of the sample and minimises the fraction of catastrophic outliers \citep{molino13}, improving the merger fraction estimation. On the other hand, our sample becomes incomplete and could be biased toward a population of either bright galaxies or galaxies with marked features in the SED (i.e., emission line galaxies or old populations with a strong $4000\AA$ break). In this section we study how the merger fraction in ALHAMBRA depends on the $\mathcal{O}$ selection and derive the optimal one to estimate the cosmic variance.

Following the methodology from spectroscopic surveys \citep[e.g.,][]{lin04,deravel09,clsj11mmvvds,clsj13ffmassiv}, if we have a population with a total number of galaxies $N_{\rm tot}$ in a given volume and we observe a random fraction $f_{\rm obs}$ of these galaxies, the merger fraction of the total population is
\begin{equation}
f_{\rm m} = f_{\rm m, obs} \times f_{\rm obs}^{-1},\label{fmp}
\end{equation}
where $f_{\rm m, obs}$ is the merger fraction of the observed sample. In ALHAMBRA we applied a selection in the parameter $\mathcal{O}$, so Eq.~(\ref{fmp}) becomes
\begin{equation}
f_{\rm m} = f_{\rm m}\,(\geq \mathcal{O}_{\rm sel}) \times \frac{N_{\rm tot}}{N\,(\geq \mathcal{O}_{\rm sel})},\label{fmal}
\end{equation}
where $N\,(\geq \mathcal{O}_{\rm sel})$ is the number of galaxies with \texttt{odds} higher than $\mathcal{O}_{\rm sel}$ (i.e., galaxies with $\mathcal{O} \geq \mathcal{O}_{\rm sel}$), $N_{\rm tot}$ is the total number of galaxies (i.e., galaxies with $\mathcal{O} \geq 0$), and $f_{\rm m}\,(\geq \mathcal{O}_{\rm sel})$ is the merger faction of those galaxies with $\mathcal{O} \geq \mathcal{O}_{\rm sel}$. Because $f_{\rm m}$ must be independent of the $\mathcal{O}$ selection, the study of $f_{\rm m}$ as a function of $\mathcal{O}_{\rm sel}$ provides the clues about the optimal \texttt{odds} selection for merger fraction studies in ALHAMBRA. We show $f_{\rm m}$ as a function of $\mathcal{O}_{\rm sel}$ for galaxies with $i \leq 22.5$ at $0.3 \leq z < 0.9$ in Fig.~\ref{fmvsodds}. We find that

\begin{itemize}
\item the merger fraction is roughly constant for $0.2 \leq \mathcal{O}_{\rm sel} \leq 0.6$. This is the expected result if the merger fraction is reliable and measured in a non biased sample. In this particular case, the $\mathcal{O}_{\rm sel} = 0.2$ (0.6) sample comprises 98\% (66\%) of the total number of galaxies with $i \leq 22.5$;

\item the merger fraction is overestimated for $\mathcal{O}_{\rm sel} \leq 0.1$. Even if only a small fraction of galaxies with poor constrains in their $z_{\rm p}$'s are included in the sample, the projection effects become important;

\item the merger fraction is overestimated for $\mathcal{O}_{\rm sel} \geq 0.7$. This behaviour at high \texttt{odds} (i.e., in samples with high quality photometric redshifts) suggests that the retained galaxies are a biased sub-sample of the general population under study.
\end{itemize}

In the analysis above we only accounted for close companions of $i \leq 22.5$ galaxies with $10h^{-1}\ {\rm kpc} \leq r_{\rm p} \leq 30h^{-1}$ kpc, but we can use other values of $r_{\rm p}^{\rm max}$ or searching over different samples. On the one hand, we repeated the study for $r_{\rm p}^{\rm max} = 40$ and 50$h^{-1}$ kpc, finding the same behaviour than for $r_{\rm p}^{\rm max} = 30 h^{-1}$ kpc (Fig.~\ref{fmvsodds}). The only differences are that the merger fraction increases with the search radius and that the $\mathcal{O}_{\rm sel} = 0.2$ point starts to deviate from the expected value (the search area increases with $r_{\rm p}^{\rm max}$ and more accurate $z_{\rm p}$'s are needed to avoid projection effects). On the other hand, we explored a wide range of $i-$band magnitude selections, from $i \leq 23$ to 20, in the three previous $r_{\rm p}^{\rm max}$ cases. We find again the same behaviour. That reinforces our arguments above and suggests $0.3 \leq \mathcal{O}_{\rm sel} \leq 0.6$ as acceptable \texttt{odds} limits to select samples for merger fraction studies in ALHAMBRA.

\begin{figure}[t]
\centering
\resizebox{\hsize}{!}{\includegraphics{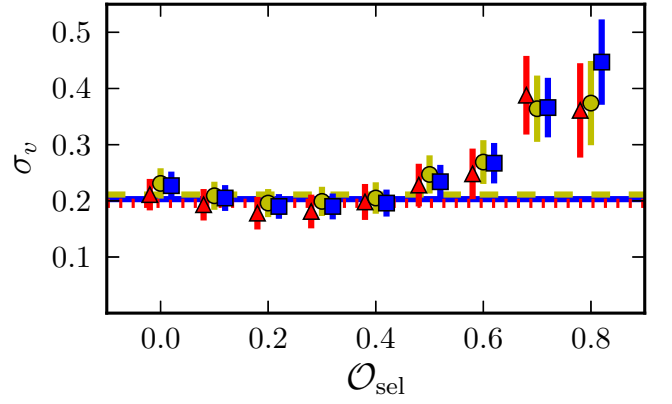}}
\caption{Cosmic variance $\sigma_v$ as a function of the \texttt{odds} selection $\mathcal{O}_{\rm sel}$ for $i \leq 22.5$ galaxies at $0.3 \leq z < 0.9$. Triangles, circles, and squares are for $r_{\rm p}^{\rm max} = 30, 40$, and 50$h^{-1}$ kpc close pairs, respectively. The dotted, dashed, and solid lines mark the average $\sigma_v$ at $0.1 \leq \mathcal{O_{\rm sel}} \leq 0.5$ for $r_{\rm p}^{\rm max} = 30, 40$, and 50$h^{-1}$ kpc close pairs, respectively. [{\it A colour version of this plot is available in the electronic edition}].}
\label{sigvvsodds}
\end{figure}

The merger fraction increases with the search radius (Fig.~\ref{fmvsodds}). However, the merger rate $R_{\rm m}$ (Sect.~\ref{secmr}) is a physical property of any population and it can not depend on $r_{\rm p}^{\rm max}$. Thus, the increase in the merger fraction with the search radius is compensated with the increase in the merger time scale \citep[e.g.,][]{deravel09,clsj11mmvvds}. This is, $R_{\rm m} \propto f_{\rm m}(r_{\rm p}^{\rm max})/T_{\rm m}(r_{\rm p}^{\rm max})$. For the same reason, the cosmic variance of the merger rate can not depend on $r_{\rm p}^{\rm max}$. In other words, the 68\% confidence interval of the merger rate, $[R_{\rm m}{\rm e}^{-\sigma_v}, R_{\rm m}{\rm e}^{\sigma_v}]$, should be independent of the search radius. Expanding the previous confidence interval we find that
\begin{eqnarray}
[R_{\rm m}{\rm e}^{-\sigma_v}, R_{\rm m}{\rm e}^{\sigma_v}] \propto \label{mrconf} \\\nonumber
[f_{\rm m}(r_{\rm p}^{\rm max})\,T^{-1}_{\rm m}(r_{\rm p}^{\rm max})\,{\rm e}^{-\sigma_v}, f_{\rm m}(r_{\rm p}^{\rm max})\,T^{-1}_{\rm m}(r_{\rm p}^{\rm max})\,{\rm e}^{\sigma_v}] = \\\nonumber
[f_{\rm m}(r_{\rm p}^{\rm max})\,{\rm e}^{-\sigma_v}, f_{\rm m}(r_{\rm p}^{\rm max})\,{\rm e}^{\sigma_v}]\,T^{-1}_{\rm m}(r_{\rm p}^{\rm max}).
\end{eqnarray}
Note that the dependence on $r_{\rm p}^{\rm max}$ is encoded in the median merger fraction and in the merger time scale. Thus, {\it the cosmic variance $\sigma_v$ of the merger fraction should not depend on the search radius}. We checked this prediction by studying the cosmic variance as a function of the search radius for $i \leq 22.5, 21.5$, and 21 galaxies with $\mathcal{O} \geq \mathcal{O}_{\rm sel} = 0.3$ at $0.3 \leq z < 0.9$. We find that $\sigma_v$ is consistent with a constant value irrespective of $r_{\rm p}^{\rm max}$ in the three populations probed, as desired (Table~\ref{sigv_rp_tab} and Fig.~\ref{sigv_vs_rp}). This supports $\sigma_v$ as a good descriptor of the cosmic variance and our methodology to measure it. In the previous analysis we have omitted the merger probability $C_{\rm m}$, which mainly depends on $r_{\rm p}^{\rm max}$ and environment (Sect.~\ref{secmr}). The merger fraction correlates with environment, so the merger probability could modify the factor ${\rm e}^{\sigma_v}$ in Eq.~(\ref{mrconf}). Because a constant $\sigma_v$ with $r_{\rm p}^{\rm max}$ is observed, the impact of $C_{\rm m}$ in the $f_{\rm m}$ to $R_{\rm m}$ translation should be similar in the range of $r_{\rm p}^{\rm max}$ explored. Detailed cosmological simulations are needed to clarify this issue.

Finally, we studied the dependence of $\sigma_v$ on the \texttt{odds} selection for $i \leq 22.5$ galaxies at $0.3 \leq z < 0.9$. Following the same arguments than before, {\it the cosmic variance should not depend on the \texttt{odds} selection}. We find that (i) $\sigma_v$ is consistent with a constant value as a function of $r_{\rm p}^{\rm max}$ for any $\mathcal{O}_{\rm sel}$, reinforcing our results above, and (ii) $\sigma_v$ is independent of the \texttt{odds} selection at $0.1 \leq \mathcal{O}_{\rm sel} \leq 0.5$ (Fig.\ref{sigvvsodds}). As for the merger fraction, we checked that different populations follow the same behaviour. We set therefore $\mathcal{O} \geq \mathcal{O}_{\rm sel} = 0.3$ as the optimal \texttt{odds} selection to measure the cosmic variance in ALHAMBRA. This selection provides excellent photometric redshifts and ensures representative samples. 

In summary, in the following we estimate the cosmic variance $\sigma_v$ from the merger fractions measured in the 48 ALHAMBRA sub-fields with $10h^{-1}$ kpc $\leq r_{\rm p} \leq 50 h^{-1}$ kpc close pairs (the $\sigma_v$ uncertainty is lower for larger search radii) and in samples with $\mathcal{O} \geq \mathcal{O}_{\rm sel} = 0.3$. That ensures reliable results in representative (i.e., non biased) samples.

\subsection{Characterisation of $\sigma_{v}$}\label{sigv}

At this stage we have set both the methodology to compute a robust cosmic variance from the observed merger fraction distribution (Sect.~\ref{ffdist}) and the optimal search radius and \texttt{odds} selection to estimate $\sigma_v$ in ALHAMBRA (Sect.~\ref{optimal}). Now we can characterise the cosmic variance as a function of the populations under study (Sects.~\ref{secn1} and \ref{secn2}), the probed cosmic volume (Sect.~\ref{secvc}), and the redshift (Sect.~\ref{secz}).

\subsubsection{Dependence on the number density of the principal sample}\label{secn1}

\begin{figure}[t]
\centering
\resizebox{\hsize}{!}{\includegraphics{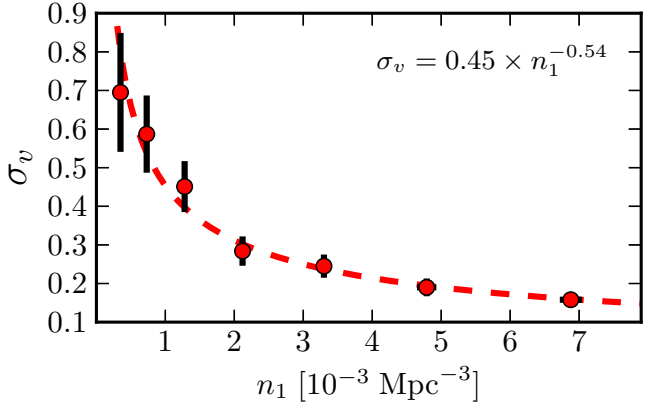}}
\caption{Cosmic variance $\sigma_v$ as a function of the number density $n_1$ of the principal population under study. Increasing the number density, the principal sample comprises $i \leq 20$, 20.5, 21, 21.5, 22, 22.5, and 23 galaxies, respectively. The probed cosmic volume is the same in all the cases, $V_c \sim 1.4 \times 10^{5}$ Mpc$^{3}$ ($0.3 \leq z < 0.9$). The dashed line is the error-weighted least-squares fit of a power-law to the data, $\sigma_{v} \propto n_1^{-0.54}$. [{\it A colour version of this plot is available in the electronic edition}].}
\label{sigv_vs_n1}
\end{figure}

In this section we explore how the cosmic variance depends on the number density $n_1$ of the principal population under study. For that, we took the same population as principal and companion sample. We study the dependence on the companion sample in Sect.~\ref{secn2}. To avoid any dependence of $\sigma_v$ on either the probed cosmic volume and $z$, and to minimise the observational errors, in this section we focus in the redshift range $0.3 \leq z < 0.9$. This range probes a cosmic volume of $V_c \sim 1.4 \times 10^{5}$ Mpc$^{3}$ in each ALHAMBRA sub-field. To explore different number densities, we measured the cosmic variance for different $i-$band selected samples, from $i \leq 20$ to $i \leq 23$ in 0.5 magnitude steps. We estimated the average number density $n_1$ in the redshift range $z_{\rm r}$ as the median number density in the 48 ALHAMBRA sub-fields, with
\begin{equation}
n_1^j\,(z_{\rm r}) = \frac{\sum_i \int_{z_{\rm min}}^{z_{\rm max}} P_i^j\, (z_i\,|\,z_{{\rm p},i},\sigma_{z_{{\rm p},i}})\, {\rm d}z_i}{V_c^j (z_{\rm r})}
\end{equation}
being the number density in the sub-field $j$ and $V_c^j$ the cosmic volume probed by it at $z_{\rm r}$. In the measurement of the number density all the galaxies were taking into account, i.e., any \texttt{odds} selection was applied ($\mathcal{O} \geq 0$). We stress that our measured number densities are unaffected by cosmic variance, and they can be used therefore to characterise $\sigma_{v}$. We report our measurements in Table~\ref{sigv_n1_tab}.

We find that the cosmic variance increases as the number density decreases (Fig.~\ref{sigv_vs_n1}), as expected by Eq.~(\ref{cosvarteofin}). The error-weighted least-squares fit of a power-law to the data is
\begin{equation}
\sigma_v\,(n_1) = (0.45 \pm 0.04) \times \bigg( \frac{n_1}{10^{-3}\ {\rm Mpc}^{-3}} \bigg)^{-0.54 \pm 0.06}.\label{sigv_n1}
\end{equation}

In this section and in the following ones we used $i-$band selected samples to characterise $\sigma_v$. We show that the results obtained with these $i-$band samples can be applied to luminosity- and stellar mass-selected samples in Sect.~\ref{secpop}.

\begin{table}
\caption{Cosmic variance $\sigma_{v}$ as a function of the principal sample's number density $n_1$}
\label{sigv_n1_tab}
\begin{center}
\begin{tabular}{lcc}
\hline\hline\noalign{\smallskip}
Principal   & $n_1$ & $\sigma_v$\\
sample      & ($10^{-3}$ Mpc$^{-3}$)    &           \\
\noalign{\smallskip}
\hline
\noalign{\smallskip}
$i \leq 23.0$ 	& $6.88 \pm 0.16$  & $0.158 \pm 0.019$\\
$i \leq 22.5$ 	& $4.79 \pm 0.14$  & $0.190 \pm 0.023$\\
$i \leq 22.0$   & $3.30 \pm 0.11$  & $0.245 \pm 0.030$\\
$i \leq 21.5$ 	& $2.12 \pm 0.07$  & $0.284 \pm 0.038$\\
$i \leq 21.0$ 	& $1.28 \pm 0.05$  & $0.451 \pm 0.066$\\
$i \leq 20.5$   & $0.73 \pm 0.03$  & $0.587 \pm 0.100$\\
$i \leq 20.0$ 	& $0.35 \pm 0.01$  & $0.695 \pm 0.154$\\
\noalign{\smallskip}
\hline
\end{tabular}
\end{center}
\end{table}

\subsubsection{Dependence on the cosmological volume}\label{secvc}

\begin{figure}[t]
\centering
\resizebox{\hsize}{!}{\includegraphics{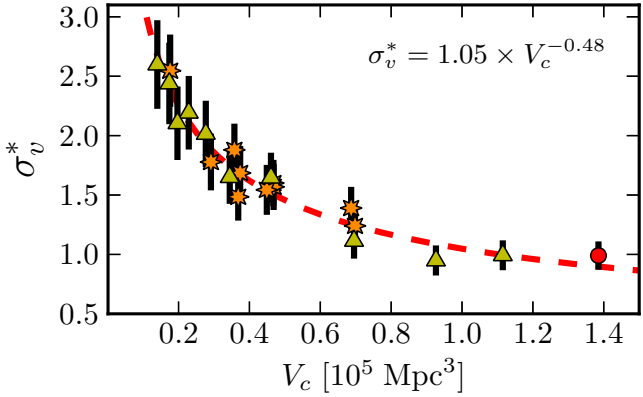}}
\caption{Normalised cosmic variance $\sigma_v^*$ as a function of the probed cosmic volume $V_c$ for galaxies with $i \leq 23$. The circle corresponds to same data as in Fig.~\ref{sigv_vs_n1}. The stars probe different redshift intervals, while triangles probe sky areas smaller than the fiducial ALHAMBRA sub-field. The dashed line is the error-weighted least-squares fit of a power-law to the data, $\sigma_{v}^{*} \propto V_c^{-0.48}$. [{\it A colour version of this plot is available in the electronic edition}].}
\label{sigv_vs_vc}
\end{figure}

In this section we explore the dependence of the cosmic variance on the cosmic volume probed by the survey. We defined $\sigma_v^{*}$ as $\sigma_v^{*} = \sigma_v / \sigma_v\,(n_1)$. This erased the dependence on the number density of the population and only volume effects were measured. We explored smaller cosmic volumes than in the previous section by studying (i) different redshift ranges over the full ALHAMBRA area (avoiding redshift ranges smaller than 0.1), and (ii) smaller areas, centred in the ALHAMBRA sub-fields, at $0.3 \leq z < 0.9$. All the cases, summarised in Table~\ref{sigv_vc_tab}, are for $i \leq 23$ galaxies. At the end, we explored an order of magnitude in volume, from $V_c \sim 0.1\times10^5$ Mpc$^{3}$ to $V_c \sim 1.4\times10^5$ Mpc$^{3}$. The power-law function that better describes the observations (Fig.~\ref{sigv_vs_vc}) is
\begin{equation}
\sigma_v^{*}\,(V_c) = (1.05 \pm 0.05) \times \bigg( \frac{V_c}{10^{5}\ {\rm Mpc}^{3}} \bigg)^{-0.48 \pm 0.05}.\label{sigv_vc}
\end{equation}

We tested the robustness of our result by fitting the two sets of data (variation in redshift and area) separately. We find $\sigma_v^{*} \propto V_c^{-0.43 \pm 0.08}$ for the redshift data, while $\sigma_v^{*} \propto V_c^{-0.48 \pm 0.05}$ for the area data.

\begin{table*}
\caption{Cosmic variance $\sigma_v$ as a function of the probed cosmic volume $V_c$}
\label{sigv_vc_tab}
\begin{center}
\begin{tabular}{lccccc}
\hline\hline\noalign{\smallskip}
Redshift      &   Effective area   & $V_c$                   & $n_1$                     & $\sigma_v$ & $\sigma^*_v$\\
range         &    (deg$^2$) & ($10^{4}$ Mpc$^{3}$)   & ($10^{-3}$ Mpc$^{-3}$)    &            & $\sigma_v/\sigma_v(n_1)$            \\
\noalign{\smallskip}
\hline
\noalign{\smallskip}
$[0.30,0.69)$   & $2.38$ &  $6.98 \pm 0.06$  &  $9.21 \pm 0.25$  & $0.169 \pm 0.025$ & $1.24 \pm 0.15$\\
$[0.69,0.90)$   & $2.38$ &  $6.87 \pm 0.06$  &  $4.69 \pm 0.16$  & $0.273 \pm 0.040$ & $1.39 \pm 0.18$\\

$[0.30,0.60)$   & $2.38$ &  $4.68 \pm 0.04$  & $10.32 \pm 0.32$  & $0.205 \pm 0.030$ & $1.60 \pm 0.19$\\
$[0.60,0.77)$   & $2.38$ &  $4.68 \pm 0.04$  &  $5.82 \pm 0.18$  & $0.274 \pm 0.042$ & $1.57 \pm 0.19$\\
$[0.77,0.90)$   & $2.38$ &  $4.49 \pm 0.04$  &  $4.17 \pm 0.19$  & $0.323 \pm 0.051$ & $1.54 \pm 0.21$\\

$[0.30,0.55)$   & $2.38$ &  $3.60 \pm 0.03$  & $11.23 \pm 0.38$  & $0.230 \pm 0.032$ & $1.88 \pm 0.22$\\
$[0.55,0.70)$   & $2.38$ &  $3.68 \pm 0.03$  &  $6.14 \pm 0.21$  & $0.252 \pm 0.041$ & $1.48 \pm 0.20$\\
$[0.70,0.82)$   & $2.38$ &  $3.74 \pm 0.03$  &  $5.26 \pm 0.21$  & $0.311 \pm 0.056$ & $1.68 \pm 0.23$\\

$[0.45,0.60)$   & $2.38$ &  $2.91 \pm 0.02$  &  $7.64 \pm 0.31$  & $0.268 \pm 0.043$ & $1.78 \pm 0.24$\\
$[0.30,0.45)$   & $2.38$ &  $1.76 \pm 0.01$  & $14.04 \pm 0.50$  & $0.276 \pm 0.051$ & $2.55 \pm 0.30$\\
\noalign{\smallskip}
\hline
\noalign{\smallskip}
$[0.30,0.90)$   & $2.38$ &  $13.85 \pm 0.11$  &  $6.88 \pm 0.16$  & $0.158 \pm 0.019$ & $0.99 \pm 0.12$\\
$[0.30,0.90)$   & $1.92$ &  $11.15 \pm 0.11$  &  $6.91 \pm 0.17$  & $0.158 \pm 0.020$ & $0.99 \pm 0.13$\\
$[0.30,0.90)$   & $1.59$ &   $9.26 \pm 0.10$  &  $7.00 \pm 0.17$  & $0.150 \pm 0.020$ & $0.95 \pm 0.13$\\
$[0.30,0.90)$   & $1.19$ &   $6.95 \pm 0.07$  &  $6.79 \pm 0.18$  & $0.179 \pm 0.024$ & $1.11 \pm 0.15$\\
$[0.30,0.90)$   & $0.79$ &   $4.61 \pm 0.05$  &  $7.06 \pm 0.20$  & $0.259 \pm 0.033$ & $1.64 \pm 0.21$\\
$[0.30,0.90)$   & $0.59$ &   $3.44 \pm 0.04$  &  $6.85 \pm 0.22$  & $0.264 \pm 0.036$ & $1.65 \pm 0.22$\\
$[0.30,0.90)$   & $0.48$ &   $2.77 \pm 0.03$  &  $6.74 \pm 0.21$  & $0.325 \pm 0.045$ & $2.01 \pm 0.28$\\
$[0.30,0.90)$   & $0.39$ &   $2.29 \pm 0.03$  &  $6.73 \pm 0.21$  & $0.354 \pm 0.050$ & $2.19 \pm 0.31$\\
$[0.30,0.90)$   & $0.34$ &   $1.97 \pm 0.04$  &  $6.72 \pm 0.24$  & $0.340 \pm 0.050$ & $2.10 \pm 0.31$\\
$[0.30,0.90)$   & $0.30$ &   $1.74 \pm 0.03$  &  $6.82 \pm 0.26$  & $0.391 \pm 0.055$ & $2.44 \pm 0.34$\\
$[0.30,0.90)$   & $0.24$ &   $1.40 \pm 0.02$  &  $6.99 \pm 0.26$  & $0.411 \pm 0.059$ & $2.60 \pm 0.37$\\
\noalign{\smallskip}
\hline
\end{tabular}
\end{center}
\end{table*}

\subsubsection{Dependence on redshift}\label{secz}
\begin{figure}[t]
\centering
\resizebox{\hsize}{!}{\includegraphics{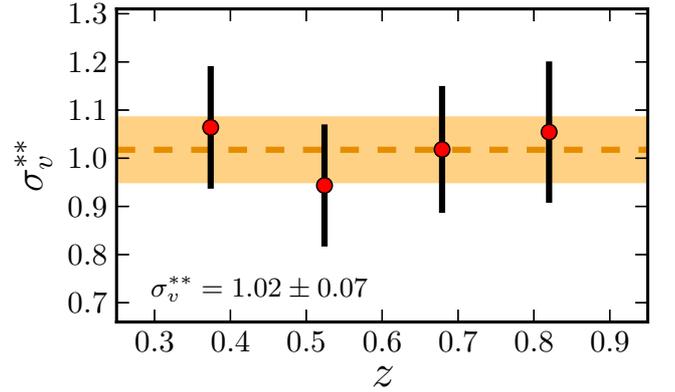}}
\caption{Normalised cosmic variance $\sigma_v^{**}$ as a function of redshift for galaxies with $i \leq 23$ (circles). The dashed line marks the error-weighted average of $\sigma_v^{**}$, $\sigma_v^{**} = 1.02 \pm 0.07$, and the coloured area shows its 68\% confidence interval. [{\it A colour version of this plot is available in the electronic edition}].}
\label{sigv_vs_z}
\end{figure}

The redshift is an expected parameter in the parametrisation the cosmic variance. However, Fig.~\ref{sigv_vs_vc} shows that the results at different redshifts are consistent with those from the wide redshift range $0.3 \leq z < 0.9$. As a consequence, the redshift dependence of the cosmic variance should be smaller than the typical error in our measurements. We tested this hypothesis by measuring $\sigma_v$ in different, non-overlapping, redshift bins. We summarise our measurements, performed for $i \leq 23$ galaxies, in Table~\ref{sigv_z_tab}. We defined $\sigma_v^{**} = \sigma_v / \sigma_v\,(n_1, V_c)$ to isolate the redshift dependence of the cosmic variance. We find that $\sigma_v^{**}$ is compatible with unity, $\sigma_v^{**} = 1.02 \pm 0.07$, and that no redshift dependence remains after accounting for the variation in $n_1$ and $V_c$ (Fig.~\ref{sigv_vs_z}). This confirms our initial hypothesis and we assume therefore $\gamma = 0$ in the following.

\begin{table*}
\caption{Cosmic variance $\sigma_v$ as a function of redshift}
\label{sigv_z_tab}
\begin{center}
\begin{tabular}{lcccccc}
\hline\hline\noalign{\smallskip}
Principal     &  Redshift      &   $\overline{z}$   & $n_1$                     &  $V_c$                  & $\sigma_v$ & $\sigma^{**}_v$\\
sample        &  range         &         & ($10^{-3}$ Mpc$^{-3}$)    &  ($10^{4}$ Mpc$^{3}$)   &  & $\sigma_v/\sigma_v(n_1,V_c)$                       \\
\noalign{\smallskip}
\hline
\noalign{\smallskip}
$i \leq 23$       &   $[0.30,0.45)$   & $0.374$ & $14.04 \pm 0.50$  & $ 1.76 \pm 0.01$  & $0.276 \pm 0.033$ & $1.06 \pm 0.13$\\
$i \leq 23$       &   $[0.45,0.60)$   & $0.524$ & $ 7.64 \pm 0.31$  & $ 2.29 \pm 0.02$  & $0.268 \pm 0.036$ & $0.94 \pm 0.13$\\
$i \leq 23$       &   $[0.60,0.75)$   & $0.679$ & $ 5.83 \pm 0.19$  & $ 4.06 \pm 0.03$  & $0.286 \pm 0.037$ & $1.02 \pm 0.13$\\
$i \leq 23$       &   $[0.75,0.90)$   & $0.820$ & $ 4.40 \pm 0.18$  & $ 5.11 \pm 0.04$  & $0.309 \pm 0.043$
& $1.05 \pm 0.15$\\
\noalign{\smallskip}
\hline
\end{tabular}
\end{center}
\end{table*}

\subsubsection{Dependence on the number density of the companion sample}\label{secn2}
As we show in Sect.~\ref{metodo}, two different populations are involved in the measurement of the merger fraction: the principal sample and the sample of companions around principal galaxies. In the previous sections the principal and the companion sample were the same, and here we explore how the number density $n_2$ of the companion sample impacts the cosmic variance. We set $i \leq 20.5$ galaxies at $0.3 \leq z < 0.9$ as principals, and varied the $i-$band selection of the companion galaxies from $i \leq 20.5$ to $i \leq 23$ in 0.5 steps. As in Sect.~\ref{secvc}, the variable $\sigma_v^{*} = \sigma_v / \sigma_v\,(n_1)$ was used.

\begin{figure}[t]
\centering
\resizebox{\hsize}{!}{\includegraphics{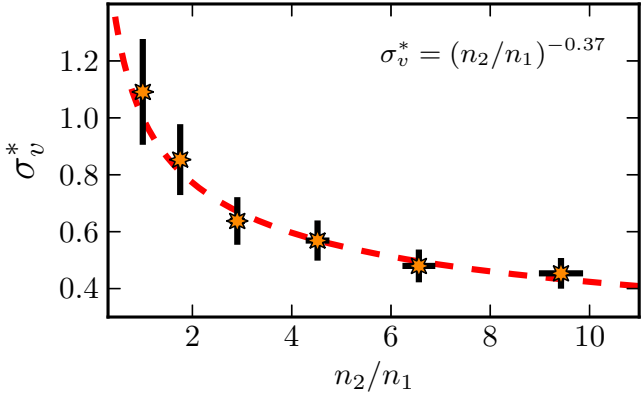}}
\caption{Normalised cosmic variance $\sigma_v^{*}$ as a function of the relative number density of the companion and the principal samples under study, $n_2/n_1$. Increasing the relative density, the companion sample comprises $i \leq 20.5$, 21, 21.5, 22, 22.5, and 23 galaxies, respectively. The red dashed line is the error-weighted least-squares fit of a power-law to the data, $\sigma_{v}^{*} = (n_2/n_1)^{-0.37}$. [{\it A colour version of this plot is available in the electronic edition}].}
\label{sigv_vs_n2}
\end{figure}

We find that the cosmic variance decreases as the number density of the companion sample increases (Table~\ref{sigv_n2_tab} and Fig.~\ref{sigv_vs_n2}). We fit the dependence with a power-law, forcing it to pass for the point $\sigma_v^{*}\,(n_1,n_1) = 1$. We find that
\begin{equation}
\sigma_v^{*}\,(n_1,n_2) = \bigg( \frac{n_2}{n_1} \bigg)^{-0.37 \pm 0.04}.\label{sigv_n2}
\end{equation}
We checked that if we leave free the intercept, it is consistent with unity, as we assumed: $\sigma_v^{*}\,(n_1,n_1) = 1.04\pm0.12$. In addition, the power-law index changes slightly, $\sigma_v^{*} \propto (n_2/n_1)^{-0.39 \pm 0.08}$.

\subsubsection{The cosmic variance in merger fraction studies bases on close pairs}
In the previous sections we have characterised the dependence of the cosmic variance $\sigma_v$ on the basic parameters involved in close pair studies (Sect.~\ref{theory}): the number density of the principal ($n_1$, Sect.~\ref{secn1}) and the companion sample ($n_2$, Sect.~\ref{secn2}), the cosmic volume under study ($V_c$, Sect.~\ref{secvc}), and the redshift (Sect.~\ref{secz}). We find that
\begin{eqnarray}
\lefteqn{\sigma_v\,(n_1,n_2,V_c) =} \nonumber\\
&& 0.48 \times \bigg( \frac{n_1}{10^{-3}\ {\rm Mpc}^{-3}} \bigg)^{-0.54} \!\!\!\! \times \bigg( \frac{V_c}{10^{5}\ {\rm Mpc}^{3}} \bigg)^{-0.48} \!\!\!\! \times \bigg( \frac{n_2}{n_1} \bigg)^{-0.37}.\label{sigv_final}
\end{eqnarray}

This is the main result of the present paper. We estimated through Monte Carlo sampling than the typical uncertainty in $\sigma_v$ from this relation is $\sim15$\%. The dependence of $\sigma_v$ on redshift should be lower than this uncertainty. In addition, $\sigma_v$ is independent of the search radius used to compute the merger fraction as we demonstrated in Sect.~\ref{optimal}.

\begin{table}
\caption{Cosmic variance $\sigma_v$ as a function of the companion sample's number density $n_2$}
\label{sigv_n2_tab}
\begin{center}
\begin{tabular}{lccc}
\hline\hline\noalign{\smallskip}
Companion & $n_2/n_1$ & $\sigma_v$ & $\sigma^{*}_v$\\
 sample &  &  &  $\sigma_v/\sigma_v(n_1)$ \\
\noalign{\smallskip}
\hline
\noalign{\smallskip}
$i \leq 20.5$ 	& $1$              & $0.587 \pm 0.100$ & $1.09 \pm 0.18$\\
$i \leq 21.0$ 	& $1.75 \pm 0.10$  & $0.459 \pm 0.067$ & $0.85 \pm 0.12$\\
$i \leq 21.5$ 	& $2.90 \pm 0.15$  & $0.343 \pm 0.045$ & $0.64 \pm 0.08$\\
$i \leq 22.0$ 	& $4.52 \pm 0.24$  & $0.306 \pm 0.038$ & $0.57 \pm 0.07$\\
$i \leq 22.5$ 	& $6.56 \pm 0.33$  & $0.258 \pm 0.031$ & $0.48 \pm 0.06$\\
$i \leq 23.0$ 	& $9.42 \pm 0.44$  & $0.244 \pm 0.029$ & $0.45 \pm 0.05$\\
\noalign{\smallskip}
\hline
\end{tabular}
\end{center}
\end{table}

\subsection{Cosmic variance in spatially random samples}\label{randsigv}
\begin{figure}[t]
\centering
\resizebox{\hsize}{!}{\includegraphics{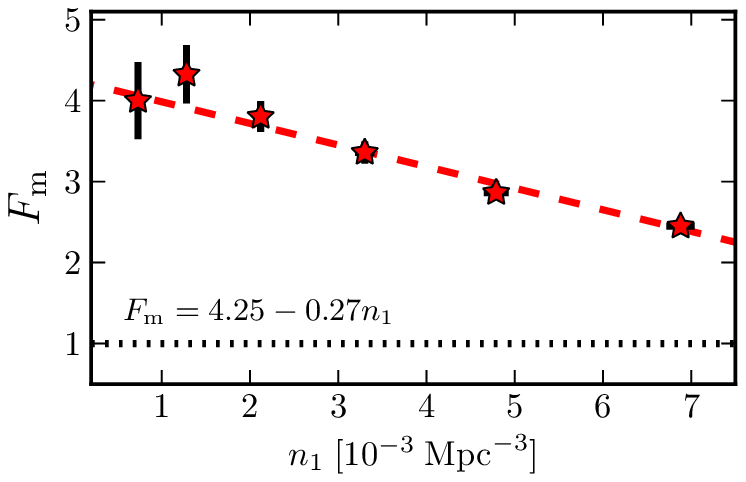}}
\resizebox{\hsize}{!}{\includegraphics{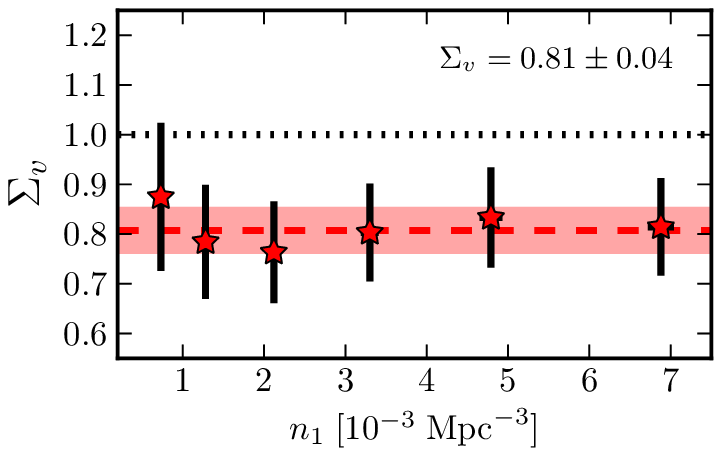}}
\caption{{\it Top panel}: merger fraction in real samples over the average merger fraction in random samples, $F_{\rm m}$, as a function of the number density $n_1$. The dotted line marks identity. The dashed line marks the best least-squares linear fit to the data, $F_{\rm m} = 4.25 - 0.27 n_1$. {\it Bottom panel}: cosmic variance in real samples over the average cosmic variance in random samples, $\Sigma_{v}$, as a function of the number density $n_1$. The dotted line marks identity. The dashed line is the error-weighted average of the data, $\Sigma_{v} = 0.81 \pm 0.04$, and the coloured area its 68\% confidence interval. [{\it A colour version of this plot is available in the electronic edition}].}
\label{randfig}
\end{figure}

In this section we further test the significance of our results by measuring both the merger fraction and the cosmic variance in samples randomly distributed in the sky plane. For this we created a set of 100 random samples, with each random sample comprising 48 random sub-samples (one per ALHAMBRA sub-field). We generated each random sub-sample by assigning a random RA and Dec to each source in the original catalogue, but retaining the original redshift of the sources. This erases the clustering signal inside each ALHAMBRA sub-field (i.e., at $\lesssim 15\arcmin$ scales), but the number density variations between sub-fields because of the clustering at scales larger than $\sim 15\arcmin$ remains. We estimated the merger fraction and the cosmic variance for each random sample at $0.3 \leq z < 0.9$ as in Sect.~\ref{secn1}, and computed the median merger fraction, $\langle f_{\rm m} \rangle$, and the median cosmic variance, $\langle \sigma_{v} \rangle$, in the set of 100 random samples to compare them with the values measured in the real samples. To facilitate this comparison, we defined the variables $F_{\rm m} = f_{\rm m}/\langle f_{\rm m} \rangle$ and $\Sigma_{v} = \sigma_{v}/\langle \sigma_{v} \rangle$. We estimated $F_{\rm m}$ and $\Sigma_{v}$ for different selections in $n_1$ following Sect.~\ref{secn1}, and we show our findings in Fig.~\ref{randfig}.

On the one hand, the merger fraction in the real samples is higher than in the random samples by a factor of three--four, $F_{\rm m} = 4.25 - 0.27 \times n_1$ (Fig.~\ref{randfig}, {\it top panel}). This reflects the clustering present in the real samples that we erased when randomised the positions of the sources in the sky, as well as the higher clustering of more luminous galaxies. This result is consistent with previous close pair studies comparing real and random samples \citep[e.g.,][]{kar07}. On the other hand, the cosmic variance measured in the random samples is higher than the cosmic variance in the real ones, $\langle \Sigma_{v} \rangle = 0.81 \pm 0.04$ (Fig.~\ref{randfig}, {\it bottom panel}). This implies that most of the variance between sub-fields is unrelated with the clustering inside these sub-fields, and that the $\sigma_v$ measured in the present paper is a real signature of the relative field-to-field variation of the merger fraction.

\subsection{Testing the independence of the 48 ALHAMBRA sub-fields}\label{sec7f}
Hitherto we have assumed that the 48 ALHAMBRA sub-fields are independent. However, only the 7 ALHAMBRA fields are really independent and correlations between adjacent sub-fields should exists. This correlations could impact our $\sigma_v$ measurements, and in this section we test the independence assumption.

\begin{figure}[t]
\centering
\resizebox{\hsize}{!}{\includegraphics{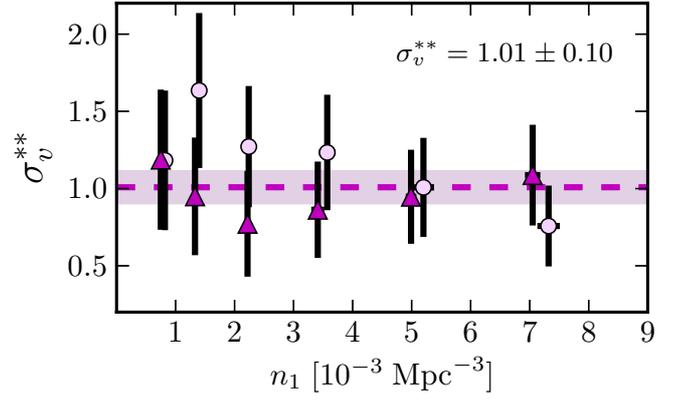}}
\caption{Normalised cosmic variance $\sigma_v^{**}$ as a function of $n_1$ at $0.3 \leq z < 0.9$ for the first (circles) and the second (triangles) group of 7 independent pointings in the ALHAMBRA survey (see text for details). The dashed line marks the error-weighted average of $\sigma_v^{**}$, $\sigma_v^{**} = 1.01\pm0.10$, and the coloured area shows its 68\% confidence interval. [{\it A colour version of this plot is available in the electronic edition}].}
\label{sigv7f}
\end{figure}

\begin{table}
\caption{Cosmic variance $\sigma_v$ measured from 7 independent pointings in the ALHAMBRA survey}
\label{sigv_7f_tab}
\begin{center}
\begin{tabular}{lccc}
\hline\hline\noalign{\smallskip}
Principal      &    $n_1$                  & $\sigma_v$  & $\sigma^{**}_v$\\
sample         &    ($10^{-3}$ Mpc$^{-3}$) &             & $\sigma_v/\sigma_v(n_1,V_c)$            \\
\noalign{\smallskip}
\hline
\noalign{\smallskip}
$i \leq 23.0$\,\tablefootmark{a}   &  $7.32 \pm 0.19$  & $0.055 \pm 0.019$ & $0.76 \pm 0.26$\\
$i \leq 22.5$\,\tablefootmark{a}   &  $5.20 \pm 0.18$  & $0.088 \pm 0.028$ & $1.01 \pm 0.32$\\
$i \leq 22.0$\,\tablefootmark{a}   &  $3.57 \pm 0.14$  & $0.132 \pm 0.040$ & $1.23 \pm 0.37$\\
$i \leq 21.5$\,\tablefootmark{a}   &  $2.24 \pm 0.09$  & $0.175 \pm 0.054$ & $1.27 \pm 0.39$\\
$i \leq 21.0$\,\tablefootmark{a}   &  $1.40 \pm 0.06$  & $0.290 \pm 0.089$ & $1.63 \pm 0.50$\\
$i \leq 20.5$\,\tablefootmark{a}   &  $0.82 \pm 0.05$  & $0.280 \pm 0.107$ & $1.18 \pm 0.45$\\
\noalign{\smallskip}
\hline
\noalign{\smallskip}
$i \leq 23.0$\,\tablefootmark{b}   &  $7.05 \pm 0.14$  & $0.080 \pm 0.024$ & $1.09 \pm 0.32$\\
$i \leq 22.5$\,\tablefootmark{b}   &  $4.99 \pm 0.09$  & $0.084 \pm 0.027$ & $0.95 \pm 0.30$\\
$i \leq 22.0$\,\tablefootmark{b}   &  $3.41 \pm 0.11$  & $0.094 \pm 0.034$ & $0.86 \pm 0.31$\\
$i \leq 21.5$\,\tablefootmark{b}   &  $2.22 \pm 0.05$  & $0.106 \pm 0.047$ & $0.77 \pm 0.34$\\
$i \leq 21.0$\,\tablefootmark{b}   &  $1.33 \pm 0.05$  & $0.172 \pm 0.069$ & $0.95 \pm 0.38$\\
$i \leq 20.5$\,\tablefootmark{b}   &  $0.75 \pm 0.02$  & $0.293 \pm 0.112$ & $1.19 \pm 0.45$\\
\noalign{\smallskip}
\hline
\end{tabular}
\end{center}
\tablefoot{\tablefoottext{a}{The 7 pointings used are: f02p01, f03p02, f04p01, f05p01, f06p01, f07p03, f08p02. The probed cosmic volume at $0.3 \leq z < 0.9$ is $V_c = (54.49 \pm 0.59)\times10^{4}$~Mpc$^{3}$.}\\
\tablefoottext{b}{The 7 pointings used are: f02p02, f03p01, f04p01, f05p01, f06p02, f07p04, f08p01. The probed cosmic volume at $0.3 \leq z < 0.9$ is $V_c = (55.24 \pm 0.50)\times10^{4}$~Mpc$^{3}$.}
}
\end{table}

We defined two groups of seven independent pointings, one per ALHAMBRA field. The first group comprises the pointings f02p01, f03p02, f04p01, f05p01, f06p01, f07p03, and f08p02; where f0? refers to the ALHAMBRA field and p0? to the pointing in the field. The second group comprises the pointings f02p02, f03p01, f04p01, f05p01, f06p02, f07p04, f08p01. Note that fields f04 and f05 have only one pointing in the current ALHAMBRA release.
Each of the previous pointings probe a cosmic volume four times higher than our fiducial sub-fields, with a median $V_c = (54.49 \pm 0.59)\times10^{4}$~Mpc$^{3}$ for the first group and $V_c = (55.24 \pm 0.50)\times10^{4}$~Mpc$^{3}$ for the second one at $0.3 \leq z < 0.9$. Then, we measured the merger fraction in the seven independent pointings of each group and we obtained $\sigma_v$ applying the MLE. We repeated this procedure for different selections, from $i \leq 23$ to $i \leq 20.5$ in 0.5 magnitude steps. Finally, we defined $\sigma_v^{**} = \sigma_v / \sigma_v\,(n_1, V_c)$, so the values of $\sigma_v^{**}$ would be dispersed around unity if the cosmic variance measured from the 7 independent areas is described well by the cosmic variance measured from the 48 sub-fields. We summarise our results in Table~\ref{sigv_7f_tab} and in Fig.~\ref{sigv7f}.

We find that the cosmic variance from the 7 independent fields nicely agree with our expectations from Eq.~(\ref{sigv_final}), with an error-weighted average of $\sigma_v^{**} = 1.01 \pm 0.10$. Thus, assume the 48 ALHAMBRA sub-fields as independent is an acceptable approximation to study $\sigma_v$. In addition, the uncertainties in $\sigma_v$ are lower by a factor of two when we use the 48 sub-fields, improving the statistical significance of our results.

\begin{table*}
\caption{Cosmic variance $\sigma_v$ of luminosity-selected samples}
\label{sigvpop_mb_tab}
\begin{center}
\begin{tabular}{lccccccc}
\hline\hline\noalign{\smallskip}
Principal &   Companion  &  Redshift    &   $n_1$   & $n_2/n_1$ &  $V_c$  &   $\sigma_v$ & $\sigma^{***}_v$\\
 sample   &     sample   &   range      & ($10^{-3}$ Mpc$^{-3}$)  &  & ($10^{4}$ Mpc$^{3}$)  &       & $\sigma_v/\sigma_v(n_1,n_2,V_c)$\\
\noalign{\smallskip}
\hline
\noalign{\smallskip}
$M_B \leq -20.5$  & $M_B \leq -20.5$  & $[0.30, 0.90)$ & $1.63 \pm 0.06$  & $1$              & $13.85 \pm 0.11$ & $0.305 \pm 0.050$ & $0.97 \pm 0.16$ \\

$M_B \leq -20.0$  & $M_B \leq -20.0$  & $[0.30, 0.90)$ & $2.95 \pm 0.08$  & $1$              & $13.85 \pm 0.11$ & $0.250 \pm 0.034$ & $1.09 \pm 0.15$ \\
$M_B \leq -20.0$  & $M_B \leq -20.0$  & $[0.30, 0.69)$ & $2.67 \pm 0.08$  & $1$              & $ 6.98 \pm 0.06$ & $0.309 \pm 0.050$ & $0.92 \pm 0.15$ \\
$M_B \leq -20.0$  & $M_B \leq -20.0$  & $[0.69, 0.90)$ & $3.37 \pm 0.11$  & $1$              & $ 6.87 \pm 0.06$ & $0.276 \pm 0.041$ & $0.92 \pm 0.14$ \\

$M_B \leq -19.5$  & $M_B \leq -19.5$  & $[0.30, 0.90)$ & $4.63 \pm 0.11$  & $1$              & $13.85 \pm 0.11$ & $0.213 \pm 0.026$ & $1.19 \pm 0.14$ \\
$M_B \leq -19.5$  & $M_B \leq -19.5$  & $[0.30, 0.60)$ & $4.12 \pm 0.15$  & $1$              & $ 4.68 \pm 0.04$ & $0.284 \pm 0.042$ & $0.88 \pm 0.13$ \\
$M_B \leq -19.5$  & $M_B \leq -19.5$  & $[0.60, 0.77)$ & $4.72 \pm 0.14$  & $1$              & $ 4.68 \pm 0.04$ & $0.288 \pm 0.038$ & $0.96 \pm 0.13$ \\
$M_B \leq -19.5$  & $M_B \leq -19.5$  & $[0.77, 0.90)$ & $5.16 \pm 0.19$  & $1$              & $ 4.49 \pm 0.04$ & $0.302 \pm 0.040$ & $1.04 \pm 0.14$ \\

$M_B \leq -19.0$  & $M_B \leq -19.0$  & $[0.30, 0.90)$ & $6.76 \pm 0.14$  & $1$              & $13.85 \pm 0.11$ & $0.165 \pm 0.020$ & $1.13 \pm 0.14$ \\
$M_B \leq -19.0$  & $M_B \leq -19.0$  & $[0.30, 0.60)$ & $6.10 \pm 0.18$  & $1$              & $ 4.68 \pm 0.04$ & $0.223 \pm 0.034$ & $0.86 \pm 0.13$ \\
$M_B \leq -19.0$  & $M_B \leq -19.0$  & $[0.60, 0.77)$ & $6.79 \pm 0.17$  & $1$              & $ 4.68 \pm 0.04$ & $0.251 \pm 0.031$ & $1.02 \pm 0.13$ \\
$M_B \leq -19.0$  & $M_B \leq -19.0$  & $[0.77, 0.90)$ & $7.25 \pm 0.28$  & $1$              & $ 4.49 \pm 0.04$ & $0.227 \pm 0.029$ & $0.94 \pm 0.12$ \\

$M_B \leq -20.5$  & $M_B \leq -20.0$  & $[0.30, 0.90)$ & $1.63 \pm 0.06$  & $1.81 \pm 0.08$  & $13.85 \pm 0.11$ & $0.262 \pm 0.035$ & $1.03 \pm 0.14$ \\
$M_B \leq -20.5$  & $M_B \leq -19.5$  & $[0.30, 0.90)$ & $1.63 \pm 0.06$  & $2.84 \pm 0.12$  & $13.85 \pm 0.11$ & $0.222 \pm 0.027$ & $1.04 \pm 0.13$ \\
$M_B \leq -20.5$  & $M_B \leq -19.0$  & $[0.30, 0.90)$ & $1.63 \pm 0.06$  & $4.12 \pm 0.12$  & $13.85 \pm 0.11$ & $0.184 \pm 0.022$ & $0.98 \pm 0.12$ \\

$M_B \leq -20.0$  & $M_B \leq -19.5$  & $[0.30, 0.90)$ & $2.95 \pm 0.06$  & $1.57 \pm 0.06$  & $13.85 \pm 0.11$ & $0.207 \pm 0.026$ & $1.07 \pm 0.13$ \\
$M_B \leq -20.0$  & $M_B \leq -19.0$  & $[0.30, 0.90)$ & $2.95 \pm 0.06$  & $2.29 \pm 0.08$  & $13.85 \pm 0.11$ & $0.171 \pm 0.020$ & $1.01 \pm 0.12$ \\

$M_B \leq -19.5$  & $M_B \leq -19.0$  & $[0.30, 0.90)$ & $4.63 \pm 0.11$  & $1.46 \pm 0.05$  & $13.85 \pm 0.11$ & $0.163 \pm 0.019$ & $1.04 \pm 0.12$ \\

\noalign{\smallskip}
\hline
\noalign{\smallskip}

$M_B \leq -20.5$  & $\mathcal{R} = 1/2$       & $[0.30, 0.90)$ & $1.63 \pm 0.06$  & $2.31 \pm 0.10$  & $13.85 \pm 0.11$ & $0.268 \pm 0.040$ & $1.16 \pm 0.17$ \\
$M_B \leq -20.5$  & $\mathcal{R} = 1/4$       & $[0.30, 0.90)$ & $1.63 \pm 0.06$  & $4.15 \pm 0.17$  & $13.85 \pm 0.11$ & $0.216 \pm 0.026$ & $1.16 \pm 0.14$ \\
$M_B \leq -20.5$  & $\mathcal{R} = 1/10$      & $[0.30, 0.60)$ & $1.41 \pm 0.05$  & $8.24 \pm 0.37$  & $ 4.68 \pm 0.04$ & $0.296 \pm 0.039$ & $1.12 \pm 0.15$ \\

$M_B \leq -20.0$  & $\mathcal{R} = 1/2$       & $[0.30, 0.90)$ & $2.95 \pm 0.08$  & $1.92 \pm 0.07$  & $13.85 \pm 0.11$ & $0.210 \pm 0.031$ & $1.16 \pm 0.17$ \\
$M_B \leq -20.0$  & $\mathcal{R} = 1/4$       & $[0.30, 0.75)$ & $2.74 \pm 0.09$  & $3.18 \pm 0.12$  & $ 8.74 \pm 0.07$ & $0.195 \pm 0.026$ & $1.01 \pm 0.13$ \\
$M_B \leq -20.0$  & $\mathcal{R} = 1/10$      & $[0.30, 0.45)$ & $2.78 \pm 0.13$  & $5.95 \pm 0.33$  & $ 1.76 \pm 0.01$ & $0.329 \pm 0.045$ & $1.00 \pm 0.14$ \\

$M_B \leq -19.5$  & $\mathcal{R} = 1/2$       & $[0.30, 0.75)$ & $4.32 \pm 0.10$  & $1.75 \pm 0.05$  & $ 8.74 \pm 0.05$ & $0.194 \pm 0.031$ & $1.03 \pm 0.16$ \\
$M_B \leq -19.5$  & $\mathcal{R} = 1/4$       & $[0.30, 0.60)$ & $4.12 \pm 0.15$  & $2.82 \pm 0.13$  & $ 4.68 \pm 0.04$ & $0.173 \pm 0.027$ & $0.79 \pm 0.12$ \\

\noalign{\smallskip}
\hline
\end{tabular}
\end{center}
\end{table*}

\begin{figure}[t]
\centering
\resizebox{\hsize}{!}{\includegraphics{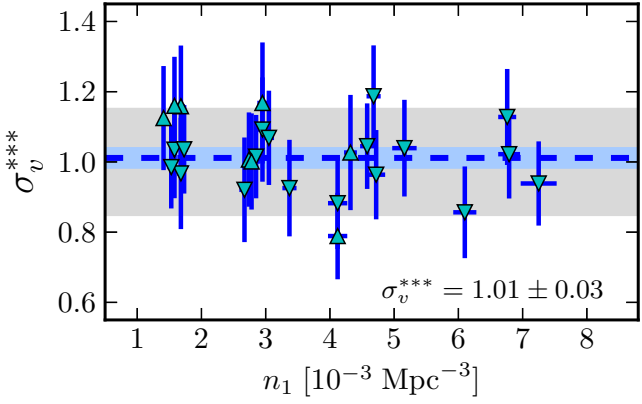}}
\caption{Normalised cosmic variance $\sigma_v^{***}$ as a function of $n_1$ for samples selected in $B$-band luminosity. The inverted triangles are those samples without a luminosity ratio imposed, and the triangles those with a luminosity ratio applied (Table~\ref{sigvpop_mb_tab}). Points at the same number density are offset when needed to avoid overlap. The dashed line marks the error-weighted average of $\sigma_v^{***}$, $\sigma_v^{***} = 1.01 \pm 0.03$. The coloured area shows its 68\% confidence interval. The grey area marks the 15\% uncertainty expected from our parametrisation of the cosmic variance. [{\it A colour version of this plot is available in the electronic edition}].}
\label{sigvpop_mb}
\end{figure}

\subsection{Expectations for luminosity- and mass-selected samples}\label{secpop}
Throughout present paper we have focused our analysis in (aparent) bright galaxies with $i \leq 23$. This ensures excellent photometric redshifts and provides reliable merger fraction measurements (Sect.~\ref{optimal}). However, one will be interested on the merger fraction of galaxies selected by their luminosity, stellar mass, colour, etc. Because the bias of the galaxies with respect to the underlying dark-matter distribution depends on the selection of the sample, our prescription to estimate $\sigma_v$ could not be valid for physically selected samples (Sect.~\ref{theory}). In this section we compare the expected cosmic variance from Eq.~(\ref{sigv_final}) with the actual cosmic variance of several luminosity- and stellar mass-selected samples to set the limits and the reliability of our suggested parametrisation.

We defined the variable $\sigma_v^{***}=\sigma_v / \sigma_v\,(n_1, n_2, V_c)$, so the values of $\sigma_v^{***}$ would be dispersed around unity if no extra dependence on the luminosity or the stellar mass exists. Throughout the present paper we imposed neither luminosity nor mass ratio constraint between the galaxies in the close pairs. However, merger fraction studies impose such constraints to study major or minor mergers. This ratio is defined as $\mathcal{R} = M_{\star,2}/M_{\star,1}$, where $M_{\star,1}$ and $M_{\star,2}$ are the stellar masses of the principal and the companion galaxy in the pair, respectively. The definition of $\mathcal{R}$ in the $B$-band luminosity $L_{B}$ case is similar. Major mergers are usually defined with $1/4 \leq \mathcal{R} \leq 1$, while minor mergers with $\mathcal{R} \leq 1/4$. We explored different $\mathcal{R}$ cases and estimated $n_2$ as the number density of the $L_{B} \geq \mathcal{R} L_{B,1}$ or the $M_{\star} \geq \mathcal{R} M_{\star,1}$ population. The properties of all the studied samples are summarised in Tables~\ref{sigvpop_mb_tab} and \ref{sigvpop_ms_tab}. The redshift range probed in each case was chosen to ensure volume-limited companion samples. We stress that the samples in Tables~\ref{sigvpop_mb_tab} and \ref{sigvpop_ms_tab} mimic typical observational selections and $\mathcal{R}$ values from the literature.

\begin{figure}[t]
\centering
\resizebox{\hsize}{!}{\includegraphics{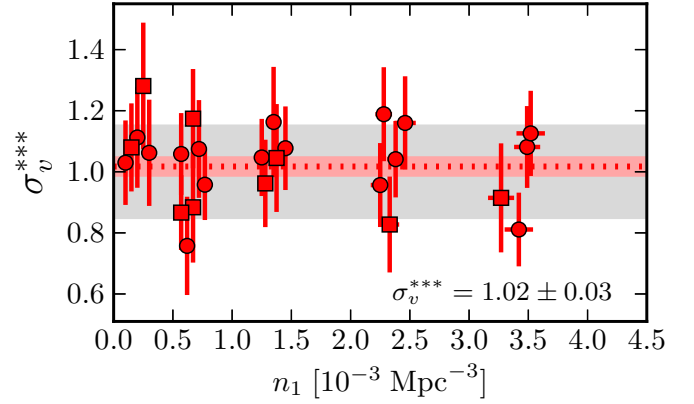}}
\caption{Normalised cosmic variance $\sigma_v^{***}$ as a function of $n_1$ for samples selected in stellar mass. The dots are those samples without a mass ratio imposed, and the squares those with a mass ratio applied (Table~\ref{sigvpop_ms_tab}). Points at the same number density are offset when needed to avoid overlap. The dashed line marks the error-weighted average of $\sigma_v^{***}$, $\sigma_v^{***} = 1.02 \pm 0.03$. The coloured area shows its 68\% confidence interval. The grey area marks the 15\% uncertainty expected from our parametrisation of the cosmic variance. [{\it A colour version of this plot is available in the electronic edition}].}
\label{sigvpop_ms}
\end{figure}

On the one hand, we find that the error-weighted average of all the luminosity-selected samples is $\sigma_v^{***} = 1.01 \pm 0.03$, compatible with unity as we expected if no (or limited) dependence on the selection exists (Fig.~\ref{sigvpop_mb}). We obtained $\sigma_v^{***} = 1.03 \pm 0.05$ from samples with the luminosity ratio $\mathcal{R}$ applied, while $\sigma_v^{***} = 1.00 \pm 0.03$ from samples without it. On the other hand, we find $\sigma_v^{***} = 1.02 \pm 0.03$ for the stellar mass-selected samples (Fig.~\ref{sigvpop_ms}). As previously, the value is compatible with unity. We obtained $\sigma_v^{***} = 0.98 \pm 0.05$ from samples with the mass ratio $\mathcal{R}$ applied, while $\sigma_v^{***} = 1.03 \pm 0.03$ from samples without it.

We conclude that our results based on $i-$band selected samples provide a good description of the cosmic variance for physically selected samples, with a limited dependence ($\lesssim 15$\%) on both the luminosity and the stellar mass selection. Thus, only $n_1$, $n_2$ and $V_c$ are needed to estimate a reliable $\sigma_v$ for merger fractions studies based on close pairs.

\begin{table*}
\caption{Cosmic variance $\sigma_v$ of stellar mass-selected samples}
\label{sigvpop_ms_tab}
\begin{center}
\begin{tabular}{lccccccc}
\hline\hline\noalign{\smallskip}
Principal &   Companion  &  Redshift    &   $n_1$   & $n_2/n_1$ &  $V_c$  &   $\sigma_v$ & $\sigma^{***}_v$\\
 sample   &     sample   &   range      & ($10^{-3}$ Mpc$^{-3}$)  &  & ($10^{4}$ Mpc$^{3}$)  &       & $\sigma_v/\sigma_v(n_1,n_2,V_c)$\\
\noalign{\smallskip}
\hline
\noalign{\smallskip}
$M_{\star} \geq 10^{10.75}\ M_{\odot}$  & $M_{\star} \geq 10^{10.75}\ M_{\odot}$  & $[0.30, 0.90)$ & $0.67 \pm 0.03$  & $1$   & $13.85 \pm 0.11$ & $0.386 \pm 0.082$ & $0.76 \pm 0.16$ \\
$M_{\star} \geq 10^{10.5}\ M_{\odot}$   & $M_{\star} \geq 10^{10.5}\ M_{\odot}$   & $[0.30, 0.90)$ & $1.35 \pm 0.05$  & $1$   & $13.85 \pm 0.11$ & $0.406 \pm 0.063$ & $1.16 \pm 0.18$ \\
$M_{\star} \geq 10^{10.25}\ M_{\odot}$  & $M_{\star} \geq 10^{10.25}\ M_{\odot}$  & $[0.30, 0.90)$ & $2.33 \pm 0.07$  & $1$   & $13.85 \pm 0.11$ & $0.309 \pm 0.040$ & $1.19 \pm 0.15$ \\
$M_{\star} \geq 10^{10.25}\ M_{\odot}$  & $M_{\star} \geq 10^{10.25}\ M_{\odot}$  & $[0.30, 0.69)$ & $2.30 \pm 0.08$  & $1$   & $ 6.98 \pm 0.06$ & $0.348 \pm 0.050$ & $0.96 \pm 0.14$ \\
$M_{\star} \geq 10^{10.25}\ M_{\odot}$  & $M_{\star} \geq 10^{10.25}\ M_{\odot}$  & $[0.69, 0.90)$ & $2.46 \pm 0.09$  & $1$   & $ 6.87 \pm 0.06$ & $0.410 \pm 0.054$ & $1.16 \pm 0.15$ \\
$M_{\star} \geq 10^{10.0}\ M_{\odot}$   & $M_{\star} \geq 10^{10.0}\ M_{\odot}$   & $[0.30, 0.90)$ & $3.49 \pm 0.11$  & $1$   & $13.85 \pm 0.11$ & $0.226 \pm 0.028$ & $1.08 \pm 0.13$ \\
$M_{\star} \geq 10^{10.0}\ M_{\odot}$   & $M_{\star} \geq 10^{10.0}\ M_{\odot}$   & $[0.30, 0.69)$ & $3.32 \pm 0.12$  & $1$   & $ 6.98 \pm 0.06$ & $0.242 \pm 0.036$ & $0.81 \pm 0.12$ \\
$M_{\star} \geq 10^{10.0}\ M_{\odot}$   & $M_{\star} \geq 10^{10.0}\ M_{\odot}$   & $[0.69, 0.90)$ & $3.62 \pm 0.12$  & $1$   & $ 6.87 \pm 0.06$ & $0.323 \pm 0.040$ & $1.13 \pm 0.14$ \\

$M_{\star} \geq 10^{11.0}\ M_{\odot}$   & $M_{\star} \geq 10^{10.5}\ M_{\odot}$   & $[0.30, 0.90)$ & $0.20 \pm 0.01$  & $ 6.75 \pm 0.40$    & $13.85 \pm 0.11$ & $0.513 \pm 0.084$ & $1.03 \pm 0.14$ \\
$M_{\star} \geq 10^{11.0}\ M_{\odot}$   & $M_{\star} \geq 10^{10.25}\ M_{\odot}$  & $[0.30, 0.90)$ & $0.20 \pm 0.01$  & $11.65 \pm 0.68$    & $13.85 \pm 0.11$ & $0.439 \pm 0.065$ & $1.11 \pm 0.16$ \\
$M_{\star} \geq 10^{11.0}\ M_{\odot}$   & $M_{\star} \geq 10^{10.0}\ M_{\odot}$   & $[0.30, 0.90)$ & $0.20 \pm 0.01$  & $17.45 \pm 1.03$    & $13.85 \pm 0.11$ & $0.350 \pm 0.047$ & $1.06 \pm 0.17$ \\

$M_{\star} \geq 10^{10.75}\ M_{\odot}$  & $M_{\star} \geq 10^{10.5}\ M_{\odot}$   & $[0.30, 0.90)$ & $0.67 \pm 0.03$  & $ 2.01 \pm 0.12$    & $13.85 \pm 0.11$ & $0.423 \pm 0.063$ & $1.08 \pm 0.16$ \\
$M_{\star} \geq 10^{10.75}\ M_{\odot}$  & $M_{\star} \geq 10^{10.25}\ M_{\odot}$  & $[0.30, 0.90)$ & $0.67 \pm 0.03$  & $ 3.48 \pm 0.19$    & $13.85 \pm 0.11$ & $0.340 \pm 0.043$ & $1.06 \pm 0.13$ \\
$M_{\star} \geq 10^{10.75}\ M_{\odot}$  & $M_{\star} \geq 10^{10.0}\ M_{\odot}$   & $[0.30, 0.90)$ & $0.67 \pm 0.03$  & $ 5.21 \pm 0.28$    & $13.85 \pm 0.11$ & $0.265 \pm 0.032$ & $0.96 \pm 0.12$ \\

$M_{\star} \geq 10^{10.5}\ M_{\odot}$   & $M_{\star} \geq 10^{10.25}\ M_{\odot}$  & $[0.30, 0.90)$ & $1.35 \pm 0.05$  & $ 1.73 \pm 0.08$    & $13.85 \pm 0.11$ & $0.307 \pm 0.039$ & $1.08 \pm 0.14$ \\
$M_{\star} \geq 10^{10.5}\ M_{\odot}$   & $M_{\star} \geq 10^{10.0}\ M_{\odot}$   & $[0.30, 0.90)$ & $1.35 \pm 0.05$  & $ 5.21 \pm 0.13$    & $13.85 \pm 0.11$ & $0.257 \pm 0.031$ & $1.05 \pm 0.13$ \\

$M_{\star} \geq 10^{10.25}\ M_{\odot}$  & $M_{\star} \geq 10^{10.0}\ M_{\odot}$   & $[0.30, 0.90)$ & $2.33 \pm 0.07$  & $ 1.50 \pm 0.07$    & $13.85 \pm 0.11$ & $0.233 \pm 0.028$ & $1.04 \pm 0.12$ \\

\noalign{\smallskip}
\hline
\noalign{\smallskip}

$M_{\star} \geq 10^{11.0}\ M_{\odot}$    & $\mathcal{R} = 1/4$   & $[0.30, 0.90)$ & $0.20 \pm 0.01$  & $ 8.75 \pm 0.53$   & $12.63 \pm 0.10$ & $0.562 \pm 0.091$ & $1.28 \pm 0.21$ \\
$M_{\star} \geq 10^{11.0}\ M_{\odot}$    & $\mathcal{R} = 1/10$  & $[0.30, 0.90)$ & $0.20 \pm 0.01$  & $17.45 \pm 1.03$   & $12.63 \pm 0.10$ & $0.367 \pm 0.049$ & $1.08 \pm 0.14$ \\

$M_{\star} \geq 10^{10.75}\ M_{\odot}$  & $\mathcal{R} = 1/2$    & $[0.30, 0.90)$ & $0.67 \pm 0.03$  & $ 2.28 \pm 0.13$   & $12.63 \pm 0.10$ & $0.332 \pm 0.068$ & $0.88 \pm 0.18$ \\
$M_{\star} \geq 10^{10.75}\ M_{\odot}$  & $\mathcal{R} = 1/4$    & $[0.30, 0.90)$ & $0.67 \pm 0.03$  & $ 4.13 \pm 0.22$   & $12.63 \pm 0.10$ & $0.354 \pm 0.049$ & $1.17 \pm 0.16$ \\
$M_{\star} \geq 10^{10.75}\ M_{\odot}$  & $\mathcal{R} = 1/10$   & $[0.30, 0.60)$ & $0.57 \pm 0.03$  & $ 8.05 \pm 0.52$   & $ 4.68 \pm 0.04$ & $0.375 \pm 0.053$ & $0.87 \pm 0.12$ \\

$M_{\star} \geq 10^{10.5}\ M_{\odot}$   & $\mathcal{R} = 1/2$    & $[0.30, 0.90)$ & $1.35 \pm 0.05$  & $ 1.86 \pm 0.09$   & $12.63 \pm 0.10$ & $0.290 \pm 0.049$ & $1.04 \pm 0.18$ \\
$M_{\star} \geq 10^{10.5}\ M_{\odot}$   & $\mathcal{R} = 1/4$    & $[0.30, 0.60)$ & $1.28 \pm 0.06$  & $ 2.95 \pm 0.18$   & $ 4.68 \pm 0.04$ & $0.390 \pm 0.058$ & $0.96 \pm 0.14$ \\

$M_{\star} \geq 10^{10.25}\ M_{\odot}$  & $\mathcal{R} = 1/2$    & $[0.30, 0.60)$ & $2.33 \pm 0.08$  & $ 1.51 \pm 0.08$   & $ 4.68 \pm 0.04$ & $0.311 \pm 0.059$ & $0.83 \pm 0.16$ \\

$M_{\star} \geq 10^{10.0}\ M_{\odot}$   & $\mathcal{R} = 1/2$    & $[0.30, 0.60)$ & $3.27 \pm 0.11$  & $ 1.50 \pm 0.07$   & $ 4.68 \pm 0.04$ & $0.287 \pm 0.056$ & $0.91 \pm 0.18$ \\
\noalign{\smallskip}
\hline
\end{tabular}
\end{center}
\end{table*}

\section{Summary and conclusions}\label{conclusions}

We use the 48 sub-fields of $\sim$180 arcmin${^2}$ in the ALHAMBRA survey (total effective area of $2.38~$deg$^2$) to estimate empirically, for the first time in the literature, the cosmic variance that affect merger fraction studies based on close pairs. We find that the distribution of the merger fraction is log-normal and we use a maximum likelihood estimator to measure the cosmic variance $\sigma_v$ unaffected by observational errors (including the Poisson shot noise term).

We find that the better parametrisation of the cosmic variance for merger fraction studies based on close pairs is (Eq.~[\ref{sigv_final}])
\begin{eqnarray}
\lefteqn{\sigma_v\,(n_1,n_2,V_c) =} \nonumber\\
&& 0.48 \times \bigg( \frac{n_1}{10^{-3}\ {\rm Mpc}^{-3}} \bigg)^{-0.54} \!\!\!\! \times \bigg( \frac{V_c}{10^{5}\ {\rm Mpc}^{3}} \bigg)^{-0.48} \!\!\!\! \times \bigg( \frac{n_2}{n_1} \bigg)^{-0.37},\nonumber
\end{eqnarray}
where $n_1$ and $n_2$ are the cosmic average number density of the principal and the companion populations under study, respectively, and $V_c$ is the cosmological volume probed by our survey in the redshift range of interest. We stress that $n_1$ and $n_2$ should be estimated from general luminosity or mass function studies and that measurements from volumes dominated by structures (e.g., clusters or voids) should be avoided. In addition, $\sigma_v$ is independent of the search radius used to compute the merger fraction. The typical uncertainty in $\sigma_v$ from our relation is $\sim15$\%. The dependence of the cosmic variance on redshift should be lower than this uncertainty. Finally, we checked that our formula provides a good estimation of $\sigma_v$ for luminosity- and mass-selected samples, as well as for close pairs with a given luminosity or mass ratio $\mathcal{R}$ between the galaxies in the pair. In the later case, $n_2$ is the average number density of those galaxies brighter or more massive than $\mathcal{R} L_{1}$ or $\mathcal{R} M_{\star,1}$, respectively.

Equation~(\ref{sigv_final}) provides the expected cosmic variance of an individual merger fraction measurement $f_{\rm m}$ at a given field and redshift range. The 68\% confidence interval of this merger fraction is $[f_{\rm m}{\rm e}^{-\sigma_{v}}, f_{\rm m}{\rm e}^{\sigma_{v}}]$. This interval is independent of the error in the measurement of $f_{\rm m}$, so both sources of uncertainty should be added to obtain an accurate description of the merger fraction error in pencil-beam surveys. If we have access to several independent fields $j$ for our study, we should combine the cosmic variance $\sigma_v^j$ of each single field with the following formula \citep[see][for details]{moster11}:
\begin{equation}
\sigma_{v, {\rm tot}}^2 = \frac{\sum_j\,({V_{c}^j}\,\sigma_{v}^j)^2}{(\sum_j V_{c}^j)^2},\label{sigv_tot}
\end{equation}
where $V_{c}^j$ is the cosmic volume probed by each single field in the redshift range of interest.

Thanks to the Eqs.~(\ref{sigv_final}) and (\ref{sigv_tot}) we can estimate the impact of cosmic variance in close pair studies from the literature. For example, \citet{bundy09} measure the major merger fraction in the two GOODS fields. We expect $\sigma_v \sim 0.42$ for massive ($M_{\star} \geq 10^{11}\ M_{\odot}$) galaxies, while $\sigma_v \sim 0.16$ for $M_{\star} \geq 10^{10}\ M_{\odot}$ galaxies. The studies of \citet{deravel09} and \citet{clsj11mmvvds} explore the merger fraction in the VVDS-Deep. We expect $\sigma_v \lesssim 0.09$ for major mergers and $\sigma_v \lesssim 0.07$ for minor mergers in this survey. \citet{lin08} explore the merger properties of $M_{B} \leq -19$ galaxies in three DEEP2 fields. We estimate $\sigma_v \sim 0.03$ for their results. Several major close pair studies have been conducted in the COSMOS field \citep[e.g.,][]{deravel11,xu12}. Focussing in mass-selected samples, we expect $\sigma_v \sim 0.17$ for massive galaxies, while $\sigma_v \sim 0.07$ for $M_{\star} \geq 10^{10}\ M_{\odot}$ galaxies. In addition, we estimate $\sigma_v \sim 0.13$ for the minor merger fractions reported by \citet{clsj12sizecos} in the COSMOS field. Regarding local merger fractions ($z \lesssim 0.1$), the expected cosmic variance in the study of \citet{depropris05} in the MGC is $\sigma_v \sim 0.03$, while $\sigma_v < 0.03$ in the study of \citet{patton00}. Finally, studies based in the full SDSS area are barely affected by cosmic variance, with $\sigma_v \lesssim 0.005$ \citep[e.g.,][]{patton08}.

Extended samples over larger sky areas are needed to constraint the subtle redshift evolution of the comic variance, as well as its dependence on the selection of the samples. Future large photometric surveys such as J-PAS\footnote{http://j-pas.org/} (Javalambre -- Physics of the accelerating universe Astrophysical Survey), that will provide excellent photometric redshifts with $\delta_z \sim 0.003$ over 8500 deg$^2$ in the northern sky, are fundamental to progress on this topic.

In the present paper we have studied in detail the intrinsic dispersion of the merger fraction measured in the 48 ALHAMBRA sub-fields. In future papers we will explore the dependence of the {\it median} merger fraction, estimated as ${\rm e}^{\mu}$, on stellar mass, colour, or morphology \citep[see][for details about the morphological classification in ALHAMBRA]{povic13}, and we will compare the ALHAMBRA measurements (both the median and the dispersion) with the expectations from cosmological simulations.

\begin{acknowledgements}
We dedicate this paper to the memory of our six IAC colleagues and friends who
met with a fatal accident in Piedra de los Cochinos, Tenerife, in February 2007,
with a special thanks to Maurizio Panniello, whose teachings of \texttt{python}
were so important for this paper.

We thank the comments and suggestions of the anonymous referee, that improved the clarity of the manuscript.

This work has mainly been funding by the FITE (Fondo de Inversiones de Teruel) and the projects AYA2006-14056 and CSD2007-00060. We also acknowledge the financial support from the Spanish grants AYA2010-15169, AYA2010-22111-C03-01 and AYA2010-22111-C03-02, from the Junta de Andalucia through TIC-114 and the Excellence Project P08-TIC-03531, and from the Generalitat Valenciana through the project Prometeo/2009/064.

A.~J.~C. (RyC-2011-08529) and C.~H. (RyC-2011-08262) are {\it Ram\'on y Cajal} fellows of the Spanish government.

\end{acknowledgements}

\bibliography{biblio}
\bibliographystyle{aa}

\appendix

\section{Maximum likelihood estimation of the cosmic variance $\sigma_{v}$}\label{mlmethod}

Maximum likelihood estimators (MLEs) have been used in a wide range of topics in astrophysics. For example, \citet{naylor06} use a MLE to fit colour-magnitude diagrams, \citet{arzner07} to improve the determination of faint X-ray spectra, \citet{makarov06} to improve distance estimates using red giant branch stars, and \citet{clsj08ml,clsj09ffgs,clsj09ffgoods,clsj10megoods} to estimate reliable merger fractions from morphological criteria. MLEs are based on the estimation of the most probable values of a set of parameters which define the probability distribution that describes an observational sample.

The general MLE operates as follows. Throughout this Appendix we denote as ${\it P}\,({\bf a}\,|\,{\bf b})$ the probability to obtain the values ${\bf a}$, given the parameters ${\bf b}$. Being ${\bf{x}}_j$ the measured values in the ALHAMBRA field $j$ and $\theta$ the parameters that we want to estimate, we may express the joined likelihood function as
\begin{equation}
L({\bf x}_j\,|\,\theta ) \equiv -\ln \big[ \prod_j {\it P}\,( {\bf x}_j\,|\,\theta) \big] = - \sum_j \ln \big[{\it P}\,({\bf x}_j\,|\,{\bf \theta})\big].\label{MLdef}
\end{equation}
If we are able to express ${\it P}\,({\bf x}_j\,|\,\theta)$ analytically, we can minimise Eq.~(\ref{MLdef}) to obtain the best estimation of the parameters $\theta$, denote as $\theta_{\rm ML}$. In our case, ${\bf x}_j$ is the observed value of the merger fraction in log-space for the ALHAMBRA sub-field $j$, ${\bf x}_j \equiv f'_{{\rm m},j} = \ln f_{{\rm m},j}$. We decided to work in log-space because that makes the problem analytic and simplifies the implementation of the method without losing mathematical rigour.

ALHAMBRA sub-fields are assumed to have a real merger fraction (not affected by observational errors) that define a Gaussian distribution in log-space,
\begin{equation}
P_{G}\,(f'_{{\rm real},j}\,|\,\mu, \sigma_v) = \frac{1}{\sqrt{2 \pi}\,\sigma_v}\,{\rm exp}\,\bigg[-\frac{(f'_{{\rm real},j} - \mu)^2}{2 \sigma_v^2}\bigg]\,.
\end{equation}
Observational errors cause the observed $f'_{{\rm m},j}$ differ from their respective real values $f'_{{\rm real},j}$. The observed $f'_{{\rm m},j}$ are assumed to be extracted for a Gaussian distribution with mean $f'_{{\rm real},j}$ and standard deviation $\sigma_{{\rm o},j}$ (the observational errors),
\begin{equation}
P_{G}\,(f'_{{\rm m},j}\,|\,f'_{{\rm real},j}, \sigma_{{\rm o},j}) = \frac{1}{\sqrt{2 \pi}\,\sigma_{{\rm o},j}}\,{\rm exp}\,\bigg[-\frac{(f'_{{\rm m},j} - f'_{{\rm real},j})^2}{2 \sigma_{{\rm o},j}^2}\bigg]\,.
\end{equation}
We assumed that the observational errors are Gaussian in log-space, i.e., that they are log-normal in observational space. This is a good approximation of the reality because we are dealing with fractions that cannot be negative and that have asymmetric confidence intervals, as shown by \citet{cameron11}. In our case, we estimated the observational errors in log-space as $\sigma_{\rm o} = \sigma_{f}/f_{\rm m}$. We checked that the values of $\sigma_{\rm o}$ derived from our jackknife errors are similar to that estimated from the Bayesian approach in \citet{cameron11}, with a difference between them $\lesssim 15\%$.

We obtained the probability ${\it P}\,({\bf x}_j\,|\,\theta)$ of each ALHAMBRA sub-field by the total probability theorem:
\begin{eqnarray}\label{tprob}
\lefteqn{P\,(f'_{{\rm m}, j}\,|\,\mu, \sigma_v, \sigma_{{\rm o},j})} \nonumber\\
&& = \int_{-\infty}^{\infty} P_G\,(f'_{{\rm real},j}\,|\,\mu, \sigma_v) \times P_G\,(f'_{{\rm m},j}\,|\,f'_{{\rm real},j}, \sigma_{{\rm o},j})\,{\rm d}f'_{{\rm real},j},\label{Pint}
\end{eqnarray}
where $f'_{{\rm m},j} = {{\bf x}_j}$ and $(\mu, \sigma_v, \sigma_{{\rm o},j}) = \theta$ in Eq.~(\ref{MLdef}). Note that the values of $\sigma_{{\rm o},j}$ are the measured uncertainties for each ALHAMBRA sub-field, so the only unknowns are the variables $\mu$ and $\sigma_v$, that we want to estimate. Note also that we integrate over the variable $f'_{{\rm real},j}$, so we are not be able to estimate the real merger fractions individually, but only the underlying Gaussian distribution that describes the sample.

The final joined likelihood function, Eq.~(\ref{MLdef}), after integrating Eq.~(\ref{Pint}), is
\begin{equation}
L\,(f'_{{\rm m},j}\,|\,\mu, \sigma_{v}, \sigma_{{\rm o},j}) = -\frac{1}{2} \sum_j \ln\,(\sigma_{v}^2 + \sigma_{{\rm o},j}^2) + \frac{(f'_{{\rm m},j} - \mu)^2}{\sigma_{v}^2 + \sigma_{{\rm o},j}^2}.
\end{equation}
With the minimisation of this function we obtain the best estimation of both $\mu$ and the cosmic variance $\sigma_{v}$, unaffected by observational errors.

In addition, we can estimate analytically the errors in the parameters above. We can obtain those via an expansion of the function $L\,(f'_{{\rm m},j}\,|\,\mu, \sigma_{v}, \sigma_{{\rm o},j})$ in Taylor's series of its variables $\theta = (\mu, \sigma_v, \sigma_{{\rm o},j})$ around the minimisation point $\theta_{\rm ML}$. The previous minimisation process made the first $L$ derivative null and we obtain
\begin{equation}
L = L(\theta_{\rm ML}) +  \frac{1}{2}(\theta - \theta_{\rm ML})^{T} H_{xy} (\theta - \theta_{\rm ML}),
\end{equation}
where $H_{xy}$ is the Hessian matrix and $T$ denotes the transpose matrix. The inverse of the Hessian matrix provides an estimate of the 68\% confidence intervals of $\mu_{\rm ML}$ and $\sigma_{\rm ML}$, as well as the covariance between them. The Hessian matrix of the joined likelihood function $L$ is defined as
\begin{equation}\label{hessian}
H_{xy} = \left(
\begin{array}{cc}
  \frac{\partial^2 L}{\partial^2 \mu} &  \frac{\partial^2 L}{\partial \mu \, \partial \sigma_{v}}\\
  \frac{\partial^2 L}{\partial \sigma_{v} \, \partial \mu} & \frac{\partial^2 L}{\partial^2 \sigma_{v}} 
\end{array}\right),
\end{equation}
with
\begin{equation}
\frac{\partial^2 L}{\partial^2 \mu} = - \sum_i \frac{1}{\sigma_{v}^2 + \sigma_{{\rm o}, j}^2},
\end{equation}
\begin{equation}
\frac{\partial^2 L}{\partial \mu \, \partial \sigma_{v}} = \frac{\partial^2 L}{\partial \sigma_{v} \, \partial \mu} = -2\sum_i \frac{\sigma_{v} (f'_{{\rm m},j} - \mu)}{(\sigma_{v}^2 + \sigma_{{\rm o}, j}^2)^2},
\end{equation}
and
\begin{equation}
\frac{\partial^2 L}{\partial^2 \sigma_{v}} = \sum_i \frac{(\sigma_{{\rm o}, j}^2 - 3\sigma_{v}^2)\times(f'_{{\rm m},j} - \mu)^2}{(\sigma_{v}^2 + \sigma_{{\rm o}, j}^2)^3} - \frac{(\sigma_{{\rm o}, j}^2 - \sigma_{v}^2)}{(\sigma_{v}^2 + \sigma_{{\rm o}, j}^2)^2}.
\end{equation}

Then, we computed the inverse of the minus Hessian, $h_{xy} = (-H_{xy})^{-1}$. Finally, and because maximum likelihood theory states that $\sigma_{\theta_{x}}^2 \leq h_{xx}$, we estimated the variances of our inferred parameters as $\sigma_{\mu}^2 = h_{11}$ and $\sigma_{\sigma_{v}}^2 = h_{22}$.

\begin{figure}[t]
\centering
\resizebox{\hsize}{!}{\includegraphics{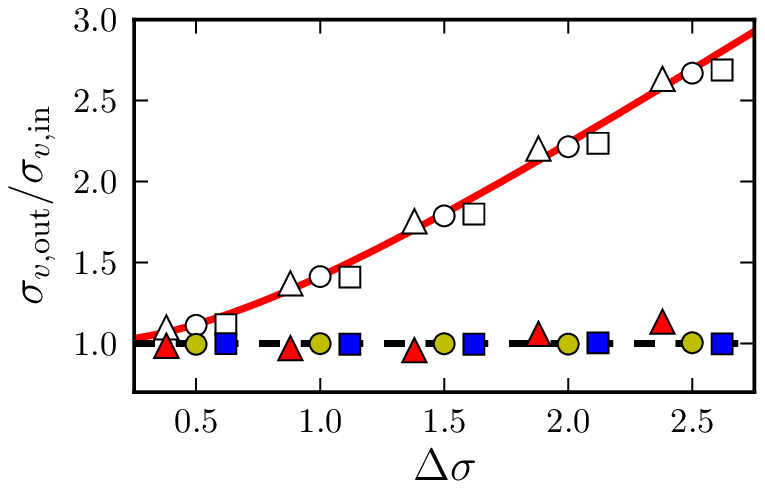}}
\resizebox{\hsize}{!}{\includegraphics{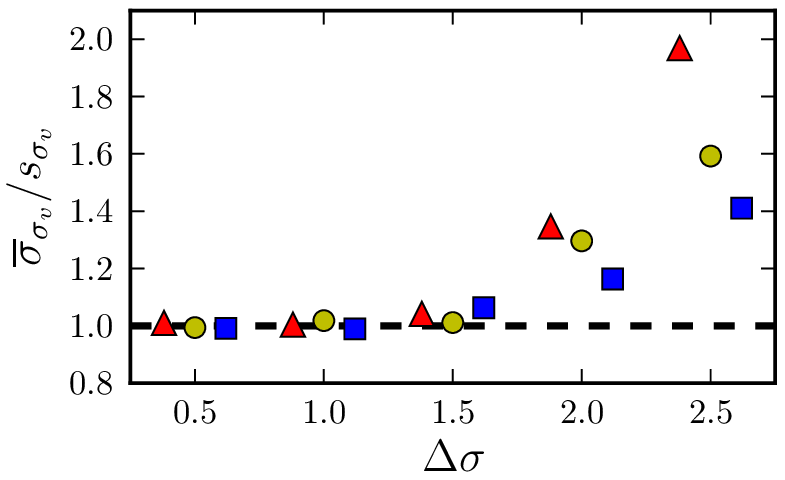}}
\caption{Recovered cosmic variance over input cosmic variance ({\it top panel}) and median $\sigma_{\sigma_v}$ over the dispersion of the recovered cosmic variance ({\it bottom panel}) as a function of $\Delta \sigma$. In both panels triangles, circles, and squares are the results from synthetic catalogues with $n = 50, 250$ and 1000, respectively. White symbols show the results from the BLS fit to the data ($\sigma_{v, {\rm BLS}}$), while those coloured show the ones from the MLE ($\sigma_{v, {\rm ML}}$). The $n = 50$ and 1000 points are shifted to avoid overlap. The dashed lines mark identity and the solid line in the {\it top panel} shows the expectation from a convolution of two Gaussians in log-space, $\sigma_{v, {\rm BLS}}/\sigma_{v, {\rm in}} = \sqrt{1 + (\Delta \sigma)^{2}}$. [{\it A colour version of this plot is available in the electronic edition}].}
\label{ml_sigv}
\end{figure}

We tested the performance and the limitations of our MLE through synthetic catalogues of merger fractions. We created several sets of 1000 synthetic catalogues, each of them composed by a number $n$ of merger fractions randomly drawn from a log-normal distribution with $\mu_{\rm in} = \log 0.05$ and $\sigma_{v, {\rm in}} = 0.2$, and affected by observational errors $\sigma_{\rm o}$. We explored the $n = 50, 250$ and $1000$ cases for the number of merger fractions, and varied the observational errors from $\sigma_{\rm o} = 0.1$ to 0.5 in 0.1 steps. That is, we explored observational errors in the measurement of the merger fraction from $\Delta \sigma \equiv \sigma_{\rm o}/\sigma_{v} = 0.5$ to 2.5 times the cosmic variance that we want to measure. We checked that the results below are similar for any value of $\sigma_{v, {\rm in}}$. We find that

\begin{enumerate}
\item The median value of the recovered $\mu$, noted $\overline{\mu}_{\rm ML}$, in each set of synthetic catalogues is similar to $\mu_{\rm in}$, with deviations lower than 0.5\% in all cases under study. However, we find that $\overline{\sigma_{v}}_{,{\rm ML}}$ for $n = 50$ catalogues overestimates $\sigma_{v, {\rm in}}$ more than 5\% at $\Delta \sigma \gtrsim 2.0$, while for $n = 1000$ we recover $\sigma_{v, {\rm in}}$ well even with $\Delta \sigma = 2.5$ (Fig.~\ref{ml_sigv}, {\it top panel}). This means that larger data sets are needed to recover the underlying distribution as the observational errors increase.

\item We also study the values recovered by a best least-squares (BLS) fit of Eq.~(\ref{Plog}) to the synthetic catalogues. We find that (i) the BLS fit recovers the right values of $\mu_{\rm in}$ as well as the MLE. That was expected, since the applied observational errors preserve the median of the initial distribution. And (ii) the BLS fit overestimates $\sigma_{v, {\rm in}}$ in all cases. The recovered values depart from the initial one as expected from a convolution of two Gaussians with variance $\sigma_{v, {\rm in}}$ and $\sigma_{\rm o}$, $\sigma_{v, {\rm BLS}}/\sigma_{v, {\rm in}} = \sqrt{1 + (\Delta \sigma)^{2}}$. The MLE performs a de-convolution of the observational errors, recovering accurately the initial cosmic variance (Fig.~\ref{ml_sigv}, {\it top panel}). 

\item The estimated variances of $\mu$ and $\sigma_{v}$ are reliable. That is, the median variances $\overline{\sigma}_{\mu}$ and $\overline{\sigma}_{\sigma_v}$ estimated by the MLE are similar to the dispersion of the recovered values, noted $s_{\mu}$ and $s_{\sigma_v}$, in each set of synthetic catalogues. The difference between both variances for $\mu$ is lower than $5$\% in all the probed cases. However, we find that $\overline{\sigma}_{\sigma_v}$ for $n = 50$ catalogues overestimates $s_{\sigma_v}$ more than 5\% at $\Delta \sigma \gtrsim 1.5$: this is the limit of the MLE to estimate reliable uncertainties with this number of data (Fig.~\ref{ml_sigv}, {\it bottom panel}). Because the estimated variance tends asymptotically to $s_{\sigma_v}$ for a large number of data, $\overline{\sigma}_{\sigma_v}$ for $n = 1000$ catalogues deviates less from the expected value than for $n = 50$ synthetic catalogues. Note that even when the estimated variance $\sigma_{\sigma_v}$ deviates from the expectations at large $\Delta \sigma$, the value of $\sigma_{v}$ is still unbiased as such large observational errors (Fig.~\ref{ml_sigv}, {\it top panel}) and we can roughly estimate $\sigma_{\sigma_v}$ through realistic synthetic catalogues as those in this Appendix.

\item The variances of the recovered parameters decreases with $n$ and increases with $\sigma_{\rm o}$. That reflects the loss of information due to the observational errors. Remark that the MLE takes these observational errors into account to estimate the parameters and their variance.
\end{enumerate}

We conclude that the MLE developed in this Appendix is not biased, provides accurate variances, and we can recover reliable uncertainties of the cosmic variance $\sigma_{v}$ in ALHAMBRA ($n = 48$) for $\Delta \sigma \lesssim 1.5$. Note that reliable values of $\sigma_{v}$ in ALHAMBRA are recovered at $\Delta \sigma \lesssim 2.0$. We checked that the average $\Delta \sigma$ in our study is 0.60 (the average observational error is $\overline{\sigma}_{\rm o} = 0.18$), and the maximum value is $\Delta \sigma = 0.85$. Thus, the results in the present paper are robust against the effect of observational errors.

\end{document}